\documentclass[journal]{IEEEtran}
\usepackage{diagbox}
\usepackage{epstopdf}
\usepackage{graphicx}
\usepackage{array}
\usepackage{multirow}
\usepackage{color}
\usepackage{amsthm,amssymb,amsfonts}
\usepackage[tight,footnotesize]{subfigure}
\usepackage{algorithm,algpseudocode}
\usepackage[cmex10]{amsmath}
\usepackage{amsthm}
\usepackage{amsmath}
\DeclareMathOperator*{\minimize}{minimize}
\DeclareMathOperator*{\maximize}{maximize}
\usepackage{color}
\usepackage{graphicx}
\usepackage{caption}
\usepackage{algorithm}
\usepackage{algorithmicx}
\usepackage{algpseudocode}
\usepackage{amsmath}
\usepackage{amssymb}
\usepackage{dsfont}
\usepackage{url}
\usepackage{CJK}
\usepackage{diagbox}
\usepackage{multirow}
\usepackage{longtable}
\usepackage{tabularx}
\usepackage{cite}
\usepackage{subfigure}

\newtheorem{lemma}{\textbf{Lemma}}

\usepackage{LTpackage}

\begin{document}
\title{Hybrid Beamforming for Millimeter Wave Systems Using the
MMSE Criterion}

\author{Tian~Lin, Jiaqi~Cong, Yu~Zhu,~\IEEEmembership{Member,~IEEE,}% <-this % stops a space
        ~Jun~Zhang,~\IEEEmembership{Senior Member,~IEEE,}% <-this % stops a space
        ~and~Khaled~B.~Letaief,~\IEEEmembership{Fellow,~IEEE}
\thanks{This work was supported by National Natural Science Foundation of China under Grant No. 61771147, and the Hong Kong Research Grants Council under Grant No. 16210216.}
\thanks{T. Lin, J. Cong, and Y. Zhu are with the Department of Communication Science and Engineering, Fudan University, Shanghai, China (e-mail: lint17@fudan.edu.cn, jqcong16@fudan.edu.cn, zhuyu@fudan.edu.cn).}
\thanks{J. Zhang is with the Department of Electronic and Information Engineering, The Hong Kong Polytechnic University (PolyU), Hung Hom, Hong Kong. Email: jun-eie.zhang@polyu.edu.hk.}
\thanks{K. B. Letaief is with the Department of Electronic and Computer Engineering, The Hong Kong University of Science and Technology, Kowloon, Hong Kong (e-mail: eekhaled@ust.hk).}}

\maketitle

\begin{abstract}
Hybrid analog and digital beamforming (HBF) has recently emerged as an attractive technique for millimeter-wave (mmWave) communication systems. It well balances the demand for sufficient beamforming gains to overcome the propagation loss and the desire to reduce the hardware cost and power consumption. In this paper, the mean square error (MSE) is chosen as the performance metric to characterize the transmission reliability. Using the minimum sum-MSE criterion, we investigate the HBF design for broadband mmWave transmissions. To overcome the difficulty of solving the multi-variable design problem, the alternating minimization method is adopted to optimize the hybrid transmit and receive beamformers alternatively. Specifically, a manifold optimization based HBF algorithm is firstly proposed, which directly handles the constant modulus constraint of the analog component. Its convergence is then proved. To reduce the computational complexity, we then propose a low-complexity general eigenvalue decomposition based HBF algorithm in the narrowband scenario and three algorithms via the eigenvalue decomposition and orthogonal matching pursuit methods in the broadband scenario. A particular innovation in our proposed alternating minimization algorithms is a carefully designed initialization method, which leads to faster convergence. {Furthermore, we extend the sum-MSE based design to that with weighted sum-MSE, which is then connected to the spectral efficiency based design.} Simulation results show that the proposed HBF algorithms achieve significant performance improvement over existing ones, and perform close to full-digital beamforming.
\end{abstract}
\IEEEpeerreviewmaketitle

% ----------------------------------------------------------------
% Section I Introduction  ***************************************
% ----------------------------------------------------------------
\section{Introduction}\label{sec:introduction}
Millimeter-wave (mmWave) communications is a key technology for 5G, which can address the bandwidth shortage problem in current mobile systems \cite{Khan:2011,Roh2014,Thomas:2014,Wang2015,Rangan2014}. The large-scale antenna array is needed to compensate for the severe path loss and penetration loss at the mmWave wavelengths \cite{Akdeniz:2014,Swindlehurst:2014}. However, the substantial increase in the number of antennas leads to non-trivial practical constraints. The traditional full-digital multiple-input and multiple-output (MIMO) beamforming which requires one dedicated radio frequency (RF) chain per antenna element is prohibitive in mmWave systems due to the unaffordable hardware cost and power consumption of a large number of antenna elements \cite{Han,Heath2014}. By separating the whole beamformer into a low-dimensional baseband digital one and a high-dimensional analog one implemented with phase shifters, the hybrid analog and digital beamforming (HBF) architecture has been shown to dramatically reduce the number of RF chains while guaranteeing a sufficient beamforming gain \cite{A.Alkhateeb2014,Heath2014,Heath2016,A2014,Bogale,Jun2018,Venugopal}.

\subsection{Related Works and Motivations}
Compared with the traditional full-digital beamforming design, in HBF, besides the difficulty of the joint optimization over the four beamforming variables (the transmit and receive analog and digital beamformers), the constant modulus constraints of the analog beamformers due to the phase shifters make the problem highly non-convex  and difficult to solve \cite{Heath2014,Lee2015,YuWei2016}. Most existing works overcome the difficulty by first decoupling the original problem into hybrid precoding and combining sub-problems and then focusing on the constant modulus constraint in solving the sub-problems. One effective and widely used approach is to regard the HBF design as a matrix factorization problem and to minimize the Euclidean distance between the hybrid beamformer with a full-digital beamformer \cite{Heath2014,Ngoc2016,ZhangJun2016}. To solve this matrix factorization problem, in \cite{Heath2014}, the authors exploited the spatial structure of the mmWave propagation channels and proposed spatially sparse precoding and combining algorithms via the orthogonal matching pursuit (OMP) method. In \cite{ZhangJun2016}, a manifold optimization (MO) based HBF algorithm, as well as some low-complexity algorithms, was proposed. Besides the matrix factorization approach, another idea for HBF design is to tackle the original problem directly. In \cite{YuWei2016,YuWei2017}, the closed-form solution of digital beamformers was first derived according to the original objective, followed by several iterative algorithms for the analog ones with the constant modulus constraint.

All  the above works, as well as most of the other previous studies, design the HBF with the objective of maximizing the spectral efficiency. By recalling the joint precoding and combining designs in conventional full-digital MIMO systems, besides  spectral efficiency, the mean square error (MSE) is another important metric \cite{Smapth:2001,Palomar:2003,Shi:2008,Jhoam}. One direct motivation to consider MSE is that a practical system is normally constrained to some particular modulation and coding scheme instead of the Gaussian code \cite{Palomar:2003}, and thus MSE is a direct performance measure to characterize the transmission reliability. { Furthermore, it has been shown that the variants of the MSE such as sum-MSE,  minmax MSE, modified MSE, weighted MSE, etc., are related to other important performance measures (e.g., signal to interference plus noise ratio (SINR) and symbol error rate) \cite{Smapth:2001,Palomar:2003,Shi:2008, Jhoam, WMMSE}. For example, it has been shown in \cite{Smapth:2001,Palomar:2003} that the MSE is related to the SINR and SER (BER) metrics in the beamforming design for the full-digital MIMO systems with multiple data streams.} Thus, it is of great interest to take MSE as an alternative optimization objective for HBF. Actually, even in some existing HBF designs with the spectral efficiency as the objective, the hybrid receive combining matrices were optimized by minimizing the MSE instead \cite{Heath2014,YuWei2016,YuWei2017,Heath2017}. Moreover, in \cite{OMP2016,Lee,Beta2008}, it was illustrated that precoding design based on the minimum MSE (MMSE) criterion can also achieve good performance in spectral efficiency.

There have been some works on the HBF design using  the MMSE  criterion for mmWave systems. In \cite{OMP2016}, the authors focused on the hybrid MMSE precoding at the transmitter side and proposed an OMP-based algorithm. To improve the system performance, in our previous work \cite{Jiaqi}, we tackled the MMSE precoding problem directly and proposed an algorithm based on the general eigen-decomposition (GEVD) method. In \cite{Ngoc2016}, the authors replaced the hybrid MMSE precoding problem by the one of factorizing the optimal full-digital MMSE precoder. In their later work \cite{Heath2017}, the hybrid MMSE combiner was further considered with a similar approach to that in \cite{Heath2014,Lee2015}, aiming at minimizing the weighted approximation gap between the hybrid combiner and a full-digital combiner. However, all of these works considered the narrowband scenario and cannot be straightforwardly extended to the broadband scenario, which is more relevant for mmWave communication systems.

\subsection{Contributions and Paper Organization}
In this paper, we investigate the joint transmit and receive HBF optimization for broadband point-to-point mmWave systems, aiming at minimizing the modified MSE \cite{Jhoam}. { Besides the aforementioned challenges in the joint optimization of the four beamforming variables and the constant modulus constraint on the analog beamformers, it is also worth noting that in the broadband scenario, yet another challenge is that the digital beamformers should be optimized for different subcarriers while the analog one is invariant for the whole frequency band. Aiming at these challenges in the MMSE based HBF design for broadband mmWave MIMO systems, the contributions in this paper can be summarized as follows.
	
\begin{itemize}
\item Instead of factorizing the optimal full-digital beamformer in the indirect HBF design approach \cite{Heath2014,Ngoc2016,ZhangJun2016}, we optimize the hybrid beamformers by directly targeting the MMSE objective for better performance. Different from the conventional MMSE based HBF designs \cite{Ngoc2016,OMP2016,Heath2017} which only considered the narrowband scenario, we propose a general HBF design approach for both the narrowband and broadband mmWave MIMO systems. In particular, we decompose the original sum-MSE minimization problem into the transmit hybrid precoding and receive combining sub-problems, and show that the two sub-problems can be unified in almost the same formulation and solved through the same procedure. The alternating minimization method is adopted to solve the overall HBF problem, for which a novel initialization method is proposed to reduce the number of iterations. Furthermore, following the approach of extending the sum-MSE minimization problem to the weighted sum-MSE minimization (WMMSE) problem and connecting it to the spectral efficiency maximization problem in the narrowband scenario \cite{Heath2017}, we show that in the broadband scenario the proposed MMSE based HBF algorithms can be generalized to the ones for maximizing the spectral efficiency.% Simulation results show that the proposed WMMSE based HBF algorithms provide better or comparable spectral efficiency than the conventional ones \cite{ZhangJun2016,YuWei2016,YuWei2017} and demonstrates the effectiveness and broad applicability of the MSE-based design approach.}

\item To deal with the constant modulus constraint in the analog beamforming optimization, we apply the manifold optimization (MO) method \cite{ZhangJun2016,MO2009}. In contrast to the application of the MO method in \cite{ZhangJun2016} for minimizing the Euclidean distance between the hybrid beamformer and the target full-digital beamformer, in this study, the MO method is applied to directly minimize the sum-MSE and the new contribution is to derive the more complicated Euclidean conjugate gradient of the sum-MSE with some skilled derivations so that the Riemannian gradient can be computed. This provides a direct approach with guaranteed convergence to solve the MMSE HBF problem instead of the indirect approach in \cite{ZhangJun2016}.

\item To avoid the high complexity in the MO-HBF algorithm, we propose several low-complexity algorithms. In the narrowband scenario, we show that the analog beamforming matrix can be optimized column-by-column with the GEVD method. In the broadband scenario, we derive both  upper  and  lower bounds of the original objective and then propose two eigen-decomposition (EVD) based HBF algorithms. Compared with the existing algorithms based on the OMP method \cite{OMP2016,Heath2017,Ngoc2016}, the proposed algorithms directly tackle the original sum-MSE objective without the restriction of the space of feasible solutions and thus result in better performance. %In addition, by combining the OMP algorithm based on the MMSE criterion for the narrowband systems \cite{OMP2016} with the one based on the maximum spectral efficiency criterion for the narrowband systems \cite{OMP2014}, we propose a new OMP-MMSE-HBF algorithm for the broadband mmWave systems.

    %{\color{red} Benefiting from the alternating optimization between the hybrid precoding and the hybrid combining, simulation results show that the proposed WMMSE based HBF algorithms provide better or comparable  spectral efficiency than the conventional ones \cite{ZhangJun2016,YuWei2016,YuWei2017}, which demonstrates the effectiveness and broad applicability of the MSE-based design approach.}

    %\item Strict convergence proof of MO based algorithm and detailed complexity analysis of all the proposed algorithms are provided in Section VI. Numerical simulations are shown in Section VII, which suggest that the MO-HBF algorithm approaches the optimal full-digital MMSE beamforming, and the proposed low-complexity algorithms achieve a good trade-off between the system performance and computational complexity.
\end{itemize}}

%The rest of the paper is organized as follows. For ease of the presentation, we start with the narrowband scenario and introduce the system model along with the HBF problem formulation based on the MMSE criterion in Section~\ref{sec:system-model-pro-fom}. To solve this MMSE HBF problem, in Section~\ref{sec:Design-Narrowband}, we present the basic idea and the optimization procedure, and propose the MO-HBF and GEVD-HBF algorithms. In Section~\ref{sec:design-broadband}, we extend the problem formulation and design procedure to the broadband scenario, and propose three HBF algorithms. { In Section \ref{sec:WMMSE}, we extend the MMSE based HBF design to the WMMSE one and propose an approach for maximizing the spectral efficiency.} We provide convergence proof for the MO-HBF algorithm and analyze the computational complexity for all the proposed HBF algorithms in Section~\ref{sec:evaluation}. We demonstrate various numerical results in Section~\ref{sec:simulation}. Finally, we conclude the paper in Section \ref{sec:conclusion}.

The rest of the paper is organized as follows. For the ease of presentation, we start with the narrowband scenario and introduce the system model along with the HBF problem formulation in Section~\ref{sec:system-model-pro-fom}. In Section~\ref{sec:Design-Narrowband}, we present the basic idea and the optimization procedure, and propose the MO-HBF and GEVD-HBF algorithms. In Section~\ref{sec:design-broadband}, we extend the problem formulation and design procedure to the broadband scenario, and propose three HBF algorithms. In Section \ref{sec:WMMSE}, we extend the MMSE based HBF design to the WMMSE one for maximizing the spectral efficiency. We discuss the convergence property and analyze the computational complexity for all the proposed HBF algorithms in Section~\ref{sec:evaluation}. We demonstrate various numerical results in Section~\ref{sec:simulation}. Finally, we conclude the paper in Section \ref{sec:conclusion}.

Throughout this paper, bold-faced upper case letters, bold-faced lower case letters, and light-faced lower case letters are used to denote matrices, column vectors, and scalar quantities, respectively. The superscripts $ (\cdot)^T$, $(\cdot)^*$, and $(\cdot)^H$ represent matrix (vector) transpose, complex conjugate, and complex conjugate transpose, respectively. $\left\| \cdot \right\|$ denotes the Euclidean norm of a vector. $\mathrm{tr}(\cdot)$,  and $\left\| \cdot \right\|_F$ denote the trace and the Frobenius norm of a matrix. $\nabla(\cdot)$ denotes the conjugate gradient of a function. $E\{\cdot\}$ denotes the expectation operator. $|.|$ denotes the absolute value or the magnitude of a complex number. $[\mathbf{A}]_{ij}$ denotes the $(i,j)$-th entry of a matrix $\mathbf{A}$.

\section{System Model and Problem Formulation}\label{sec:system-model-pro-fom}
\subsection{System Model}\label{subsec:system-model}
\begin{figure*}[!t]
\centering
\includegraphics[width=4.2in]{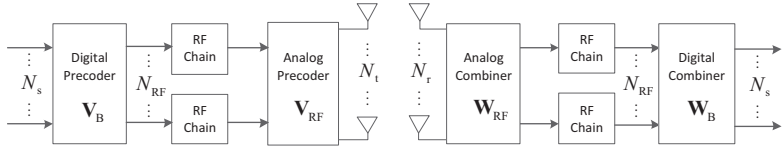}%or 8.5cm
\caption{Diagram of a point-to-point narrowband mmWave MIMO system with HBF.}
\label{fig:sys-narrowband}
\vspace{-0.8cm}
\end{figure*}

For the ease of presentation, we first consider a point-to-point narrowband mmWave MIMO system with HBF as in Fig. \ref{fig:sys-narrowband}, where $N_\mathrm{s}$ data streams are sent and collected by $N_\mathrm{t}$ transmit antennas and $N_\mathrm{r}$ receive antennas, respectively. Both the transmitter and  receiver are equipped with $N_\mathrm{RF}$ RF chains, where $\mathrm{min}(N_\mathrm{r},N_\mathrm{t}) \gg  N_\mathrm{RF}$. The original $N_\mathrm{s}\times1$ symbol vector, denoted by $\mathbf{s}$ with $E\{\mathbf{s}\mathbf{s}^H\}=\mathbf{I}_{N_\mathrm{s}}$, is firstly precoded through an $N_\mathrm{RF}\times N_\mathrm{s}$ digital beamforming matrix ${\mathbf{V}}_\mathrm{B}$, and then an $N_\mathrm{t}\times N_\mathrm{RF}$ analog beamforming matrix ${\mathbf{V}}_\mathrm{RF}$ which is implemented in the analog circuitry using phase shifters. From the equivalent baseband representation point of view, the precoded signal vector at the transmit antenna array can be represented as $\mathbf{x} ={\mathbf{V}}_\mathrm{RF} {\mathbf{V}}_\mathrm{B}\mathbf{s}$. Without loss of generality, the normalized transmit power constraint is set to $\mathrm{tr}({\mathbf{V}}_\mathrm{RF}{{\mathbf{V}}_\mathrm{B}} {\mathbf{V}}_\mathrm{B}^H{\mathbf{V}}_\mathrm{RF}^H)\le1$.

Similar to that in \cite{Heath2014,YuWei2016}, the mmWave propagation channel is characterized by a geometry-based channel model with $N_\mathrm{C}$ clusters and $N_\mathrm{R}$ rays within each cluster. Considering the mmWave system with a half-wave spaced uniform linear array (ULA) at both the transmitter and the receiver, the ${N_\mathrm{r}\times N_\mathrm{t}}$ channel matrix $\mathbf{H}$ can be represented as
\begin{equation}\label{eqn:Hmodel-narrow}
\mathbf{H} = \sqrt {\frac{{{N_\mathrm{t}}{N_\mathrm{r}}}}{{{N_\mathrm{C}}{N_{\mathrm{R}}}}}} \sum\limits_{i = 1}^{{N_\mathrm{C}}} {\sum\limits_{j = 1}^{{N_{\mathrm{R}}}} {{\alpha_{ij}}} } {\mathbf{a}_\mathrm{r}}(\theta _{ij}^\mathrm{r}){\mathbf{a}_\mathrm{t}}{(\theta_{ij}^\mathrm{t})}^H,
\end{equation}
where $\alpha_{ij}$ denotes the complex gain of the $j$th ray in the $i$th propagation cluster, and  $\mathbf{a}_\mathrm{r}(\theta_{ij}^\mathrm{r})=\frac{1}{\sqrt{N_\mathrm{r}}}\big[1 \ e^{\mathrm{j}\pi\sin{\theta^\mathrm{r}_{ij}}} \ \ldots \ e^{\mathrm{j}\pi(N_\mathrm{r}-1)\sin{\theta^\mathrm{r}_{ij}}}\big]^T$ and $\mathbf{a}_\mathrm{t}{(\theta_{ij}^\mathrm{t})}=\frac{1}{\sqrt{N_\mathrm{t}}}\big[1 \ e^{\mathrm{j}\pi\sin{\theta^\mathrm{t}_{ij}}} \ \ldots \ e^{\mathrm{j}\pi(N_\mathrm{t}-1)\sin{\theta^\mathrm{t}_{ij}}}\big]^T$ denote the normalized responses of the transmit and receive antenna arrays to the $j$th ray in the $i$th cluster, respectively, where $\theta_{ij}^\mathrm{r}$ and $\theta_{ij}^\mathrm{t}$ denote the angles of arrival and departure.

With a similar HBF at the receiver, i.e., an $N_\mathrm{r}\times N_\mathrm{RF}$ analog combiner ${\mathbf{W}}_\mathrm{RF}$ followed by an $N_\mathrm{RF}\times N_\mathrm{s}$ digital baseband combiner ${\mathbf{W}}_\mathrm{B}$, we finally have the processed signal as
\begin{equation}\label{eqn:y-tilde}
{\mathbf{y}}={\mathbf{W}}_\mathrm{B}^H{\mathbf{W}}_\mathrm{RF}^H\mathbf{H} {\mathbf{V}}_\mathrm{RF}{{\mathbf{V}}_\mathrm{B}}\mathbf{s}+ {\mathbf{W}}_\mathrm{B}^H{\mathbf{W}}_\mathrm{RF}^H\mathbf{u},
\end{equation}
where $\mathbf{u}$ denotes the additive noise vector at the $N_\mathrm{r}$ receive antennas satisfying the complex circularly symmetric Gaussian distribution with zero mean and covariance matrix ${\sigma^2}\mathbf{I}_{N_\mathrm{r}}$, i.e., $\mathbf{u} \sim \mathcal{CN}(0,{\sigma^2}\mathbf{I}_{N_\mathrm{r}})$. Similar to existing works on the HBF design (e.g. \cite{A.Alkhateeb2014,ZhangJun2016,YuWei2016,YuWei2017}), in this paper, it is assumed that perfect channel state information (CSI) is available at both the transmitter and receiver and that there is perfect synchronization between them.

\subsection{Problem Formulation}\label{subsec:pro-for}
In this work, we take the modified MSE  \cite{Jhoam} as the performance measure and optimization objective for the joint transmit and receive HBF design, which is defined as
\begin{equation}\label{eqn:def-modified-MSE}
\mathrm{MSE} \triangleq  E\{\|\beta^{-1}\mathbf{y}-\mathbf{s}\|^2\},
\end{equation}
where $\beta$ is a scaling factor to be jointly optimized with the hybrid beamformers. By substituting (\ref{eqn:y-tilde}) into (\ref{eqn:def-modified-MSE}) and after some mathematical manipulations, we have
\begin{equation}\label{eqn:MSE-definition}
\begin{aligned}
\mathrm{MSE} &= E\{\|\beta^{-1}({{\mathbf{W}}}^H\mathbf{H}{{\mathbf{V}}}\mathbf{s}+{{\mathbf{W}}}^H\mathbf{u})-\mathbf{s}\|^2\}\\
&=\mathrm{tr} 	(\beta^{-2}{\mathbf{W}}^H\mathbf{H}{\mathbf{V}}{\mathbf{V}}^H\mathbf{H}^H{\mathbf{W}} - \beta^{-1}{\mathbf{W}}^H\mathbf{H}{\mathbf{V}} \\ &- \beta^{-1}{\mathbf{V}}^H\mathbf{H}^H{\mathbf{W}}  + {\sigma^2}\beta^{-2}{\mathbf{W}}^H{\mathbf{W}}+\mathbf{I}_{N_\mathrm{s}} ),
\end{aligned}
\end{equation}
where ${{\mathbf{W}}} \triangleq {\mathbf{W}}_{\mathrm{RF}} {\mathbf{W}}_{\mathrm{B}}$, ${\mathbf{V}} \triangleq {\mathbf{V}}_{\mathrm{RF}} {\mathbf{V}}_{\mathrm{B}}$ are defined as the overall hybrid transmit and receive beamformers, respectively. Notice that since the analog beamformers are assumed to be implemented with phase shifters which only adjust the phases of the input signals, the elements of analog beamformers should satisfy the constant modulus constraint, namely $|[{\mathbf{V}}_{\mathrm{RF}}]_{ij}|=1$ for $i=1,\ldots,N_\mathrm{t}$ and $j=1,\ldots,N_\mathrm{RF}$, and $|[{\mathbf{W}}_{\mathrm{RF}}]_{ml}|=1$ for $m =1,\ldots,N_\mathrm{r}$ and $l=1,\ldots,N_\mathrm{RF}$. With the derived MSE expression in (\ref{eqn:MSE-definition}), the transmit power constraint and the constant modulus constraint of the phase shifters, the HBF optimization problem in the narrowband scenario can be formulated as
\begin{equation}\label{pro:opt-total}
\begin{array}{cl}
\displaystyle{\minimize_{{\mathbf{V}}_\mathrm{RF},{{\mathbf{V}}_\mathrm{B}}, {\mathbf{W}}_\mathrm{RF},{\mathbf{W}}_\mathrm{B},\beta}} & \mathrm{MSE} \\
\mathrm{subject \; to} & \|{\mathbf{V}}\|_F^2\leq 1;\quad |[{\mathbf{V}}_{\mathrm{RF}}] _{ij}|^2=1, \forall i,j;  \\ &|[{\mathbf{W}}_{\mathrm{RF}}] _{ml}|^2=1, \forall m,l.
\end{array}
\end{equation}
It is worth noting that there are mainly three reasons or advantages for introducing the scaling factor $\beta$ and taking the modified MSE as the objective function. First, as the joint transmit and receive HBF problem will be decoupled into the hybrid precoding and combining sub-problems, adjusting $\beta$ achieves a better performance for the precoding optimization by considering the noise effect (which is also referred to as the transmit Wiener filter) \cite{Jhoam}. Second, $\beta$ is also helpful in dealing with the total transmit power constraint and thus simplifies the precoding optimization procedure \cite{Lee,Beta2008}. Finally, by introducing $\beta$, the hybrid precoding and combining sub-problems can be unified and solved in the same way aiming at the same  modified MSE objective. These advantages will be elaborated in more details in the following sections.

% ----------------------------------------------------------------
% Section III Hybrid Beamforming Design Based On MMSE Criterion  ***************************************
% ----------------------------------------------------------------
\section{Hybrid Beamforming Design Based on The MMSE Criterion}\label{sec:Design-Narrowband}
Since the HBF problem in (\ref{pro:opt-total}) involves a joint optimization over five variables, along with non-convex constraints, it is unlikely to find the optimal solution. A sub-optimal but efficient way to overcome the difficulties is to separate the original problem into two sub-problems corresponding to the optimization for the hybrid transmit precoder and receive combiner, respectively, and solve each independently  \cite{Heath2014,YuWei2016,YuWei2017,Heath2017}. Taking this approach, we propose several HBF algorithms in the following two subsections. Finally, we develop the whole alternating minimization algorithm for the HBF optimization based on the MMSE criterion.

%Since the HBF problem in (\ref{pro:opt-total}) involves a joint optimization over five variables, along with non-convex constraints, it is unlikely to find the optimal solution. A sub-optimal but efficient way to overcome the difficulties is to separate the original problem into two sub-problems corresponding to the optimization for the hybrid transmit precoder and receive combiner, respectively, and solve each independently  \cite{Heath2014,YuWei2016,YuWei2017,Heath2017}. Taking this approach, we propose several HBF algorithms in the following two subsections. Finally, we develop the whole alternating minimization algorithm for the HBF optimization based on the MMSE criterion.

\subsection{Hybrid Transmit  Design}\label{subsec:precoding}
This section focuses on the hybrid precoder design (including $\beta$) in (\ref{pro:opt-total}) by fixing the receive combining matrices ${\mathbf{W}}_\mathrm{B}$ and $\mathbf{W}_\mathrm{RF}$. As shown in \cite{Jhoam,OMP2016,Beta2008}, the original precoder $\mathbf{V}_\mathrm{B}$ can be separated as $\mathbf{V}_\mathrm{B}=\beta\mathbf{V}_\mathrm{U}$, where $\mathbf{V}_\mathrm{U}$ is an unnormalized baseband precoder. With this separation, the precoder optimization problem can be formulated as
\begin{equation}\label{pro:opt-precoding}
\begin{array}{cl}
\displaystyle{\minimize_{\mathbf{V}_\mathrm{RF},\mathbf{V}_\mathrm{U},\beta}} & \mathrm{tr} 	(\mathbf{H}_\mathrm{1}^H\mathbf{V}_\mathrm{RF}{{\mathbf{V}}_\mathrm{U}}{{\mathbf{V}}^H_\mathrm{U}} \mathbf{V}_\mathrm{RF}^H\mathbf{H}_\mathrm{1} - \mathbf{H}_\mathrm{1}^H\mathbf{V}_\mathrm{RF}{{\mathbf{V}}_\mathrm{U}} \\  & -{{\mathbf{V}}^H_\mathrm{U}}\mathbf{V}_\mathrm{RF}^H \mathbf{H}_\mathrm{1}  + {\sigma^2}\beta^{-2} {\mathbf{W}}^H{\mathbf{W}} +\mathbf{I}_{N_\mathrm{s}} )\\ \\
\mathrm{subject \; to} & \mathrm{tr}(\mathbf{V}_\mathrm{RF}{{\mathbf{V}}_\mathrm{U}}{{\mathbf{V}}^H_\mathrm{U}}\mathbf{V}_\mathrm{RF}^H) \le \beta^{-2}; \\  &|[\mathbf{V}_{\mathrm{RF}}]_{ij}|=1, \forall i,j,
\end{array}
\end{equation}
where $\mathbf{H}_\mathrm{1}\triangleq\mathbf{H}^H \mathbf{W}_\mathrm{RF}\mathbf{W}_\mathrm{B}$ denotes the equivalent channel of the concatenation of the air interface channel and the hybrid receive combiner. Our optimization approach is to first derive the optimal digital precoding matrix ${{\mathbf{V}}_\mathrm{U}}$ and the scaling factor $\beta$ by fixing ${\mathbf{V}}_\mathrm{RF}$,  then derive the resulting objective as a function of ${\mathbf{V}}_\mathrm{RF}$, and finally optimize ${\mathbf{V}}_\mathrm{RF}$ by further minimizing the objective with the constant modulus constraint. Due to the transmit power constraint, it can be proved by contradiction that the optimal solution must be achieved with the maximum total transmit power, i.e., the optimal $\beta$ is given by
\begin{equation}\label{eqn:beta}
\beta =\left(\mathrm{tr}\left(\mathbf{V}_\mathrm{RF} {{\mathbf{V}}_\mathrm{U}} {{\mathbf{V}}_\mathrm{U}}^H\mathbf{V}_\mathrm{RF}^H\right)\right)^{-\frac{1}{2}}.
\end{equation}
Then according to the Karush-Kuhn-Tucker (KKT) conditions, the closed-form solution of the optimal $\mathbf{V}_\mathrm{U}$ is given by
\begin{equation}\label{eqn:Lag-VD-result}
{{\mathbf{V}}_\mathrm{U}} =(\mathbf{V}_\mathrm{RF}^H\mathbf{H}_\mathrm{1}\mathbf{H}_\mathrm{1}^H\mathbf{V}_\mathrm{RF}+{\sigma^2}w \mathbf{V}_\mathrm{RF}^H\mathbf{V}_\mathrm{RF})^{-1}\mathbf{V}_\mathrm{RF}^H \mathbf{H}_\mathrm{1},
\end{equation}
where $w \triangleq\mathrm{tr}( {\mathbf{W}}^H{\mathbf{W}})$ is defined for notational brevity. Substituting the optimal ${{\mathbf{V}}_\mathrm{U}}$ and $\beta$ into (\ref{pro:opt-precoding}) and after some mathematical derivation, the resulting MSE is given by\footnote{{Note that the above derivations benefit from the introduction of $\beta$. To show this, it can be checked that if we remove $\beta$ from (\ref{pro:opt-precoding}) (or just set $\beta=1$ in (\ref{pro:opt-precoding})), it is highly challenging to get a closed-form expression of $\VB$ via the KKT conditions and further get a closed-form expression of the MSE as a function of $\VRF$ for the optimization of the analog precoder}. }
\begin{equation}\label{eqn:MSE-as-fun-VRF}
J(\mathbf{V}_\mathrm{RF})\triangleq\mathrm{tr}({(\mathbf{I}_{\mathrm{N}_{\mathrm{s}}} + \frac{1}{{\sigma^2w}}\mathbf{H}_\mathrm{1}^H\mathbf{V}_\mathrm{RF}{\left(\mathbf{V}_\mathrm{RF}^H {\mathbf{V}_\mathrm{RF}}\right)^{-1}}\mathbf{V}_\mathrm{RF}^H\mathbf{H}_\mathrm{1})^{-1}}).
\end{equation}
The optimizing problem in (\ref{pro:opt-precoding}) is now reduced to the following one for the optimization of $\mathbf{V}_\mathrm{RF}$
\begin{equation}\label{pro:opt-V-RF}
\begin{array}{cl}
\displaystyle{\minimize_{\mathbf{V}_{\mathrm{RF}}}} & J(\mathbf{V}_\mathrm{RF})\\
\mathrm{subject \; to} & |[\mathbf{V}_{\mathrm{RF}}]_{ij}|=1, \forall i,j.
\end{array}
\end{equation}
Here we propose two algorithms for optimizing the analog precoding matrix $\mathbf{V}_{\mathrm{RF}}$ with the constant modulus constraint, which are based on the MO and GEVD methods, respectively.

\subsubsection{Analog Precoder Design Based on the MO Method}\label{subsubsec:manifold-narrow}
To deal with the constant modulus constraint, the MO method \cite{ZhangJun2016}\cite{MO2009} can be applied to obtain a local optimal $\mathbf{V}_{\mathrm{RF}}$. The basic idea is to define a Riemannian manifold for $\mathbf{V}_{\mathrm{RF}}$ with the consideration of the constant modulus constraint, and iteratively update this optimization variable on the direction of the Riemannian gradient (i.e., a projection of the Euclidean conjugate gradient onto the tangent space of a point on the Riemannian manifold) in a similar way to that in the conventional Euclidean gradient descent algorithm (the details can be referred to \cite{ZhangJun2016}). However, the application of the MO method is not straightforward, and the most difficult part is the derivation of the conjugate gradient in the Euclidean space, in order to obtain the associated Riemannian gradient. It should be mentioned that for the scalar function $J(\mathbf{V}_{\mathrm{RF}})$ associated with a complex-valued variable $\mathbf{V}_{\mathrm{RF}}$, the conjugate gradient \cite{AH} is defined as $\nabla J(\mathbf{V}_\mathrm{RF})=\frac{{\partial J(\mathbf{V}_\mathrm{RF})}}{{\partial \mathbf{V}_\mathrm{RF}^*}}$.
%\begin{equation}
%\nabla J(\mathbf{V}_\mathrm{RF})=\frac{{\partial J(\mathbf{V}_\mathrm{RF})}}{{\partial \mathbf{V}_\mathrm{RF}^*}}.
%\end{equation}
By defining $\mathbf{P}\triangleq\mathbf{I}_{\mathrm{N}_{\mathrm{s}}}+$ $ \frac{1}{{\sigma^2w}}\mathbf{H}_\mathrm{1}^H\mathbf{V}_\mathrm{RF} {\left(\mathbf{V}_\mathrm{RF}^H {\mathbf{V}_\mathrm{RF}}\right)^{ - 1}}\mathbf{V}_\mathrm{RF}^H\mathbf{H}_\mathrm{1}$  in (\ref{eqn:MSE-as-fun-VRF}) for notational brevity, we have the following lemma for the conjugate gradient.
\begin{lemma}\label{lemma:MO}
The conjugate gradient of the function $J(\mathbf{V}_{\mathrm{RF}})$ with respect to $\mathbf{V}_{\mathrm{RF}}$ is given by
\begin{equation}\label{eqn:gradient-J}
\begin{aligned}
\nabla J(\mathbf{V}_\mathrm{RF})= &\frac{1}{{\sigma^2w}} \left(\mathbf{V}_\mathrm{RF} {\left(\mathbf{V}_\mathrm{RF}^H{\mathbf{V}_\mathrm{RF}}\right)^{-1}} \mathbf{V}_\mathrm{RF}^H -\mathbf{I}_{N_\mathrm{t}}\right)  \\&\times\mathbf{H}_\mathrm{1}\mathbf{P}^{-2} \mathbf{H}_\mathrm{1}^H \mathbf{V}_\mathrm{RF} {\left(\mathbf{V}_\mathrm{RF}^H{\mathbf{V}_\mathrm{RF}}\right)^{-1}}.
\end{aligned}
\end{equation}
\end{lemma}

\textit{Proof}: According to some basic differentiation rules for complex-value matrices \cite{AH}, the differential of $J(\mathbf{V}_\mathrm{RF})$ can be expressed as
\begin{equation}\label{eqn:diff-J-definition}
\begin{aligned}
\mathrm{d}(J(\mathbf{V}_\mathrm{RF})) &= \mathrm{tr}\left(\left(\nabla J(\mathbf{V}_\mathrm{RF})\right)^T\mathrm{d}(\mathbf{V}^*_\mathrm{RF})\right) \\&= \mathrm{tr} \left(\nabla J(\mathbf{V}_\mathrm{RF})\mathrm{d}(\mathbf{V}^H_\mathrm{RF})\right),
\end{aligned}
\end{equation}
where $\mathrm{d}(\cdot)$ denotes the differential with respect to $\mathbf{V}_\mathrm{RF}^*$ while  taking $\mathbf{V}_\mathrm{RF}$  as a constant matrix during the derivation of the conjugate gradient $  \nabla J(\mathbf{V}_\mathrm{RF})$. The second equality in (\ref{eqn:diff-J-definition}) holds due to the properties of $\mathrm{tr}(\mathbf{A}^T) =\mathrm{tr}(\mathbf{A})$ and $\mathrm{tr}(\mathbf{A}\mathbf{B})= \mathrm{tr}(\mathbf{B}\mathbf{A})$.

On the other hand, we can directly compute $\mathrm{d}(J(\mathbf{V}_\mathrm{RF}))$ from (\ref{eqn:MSE-as-fun-VRF}).  According to some differentiation rules for differentiating a matrix's trace and inverse, we express $\mathrm{d}(J(\mathbf{V}_\mathrm{RF}))$ as
\begin{equation}\label{eqn:diff-J-derive-1}
\mathrm{d}(J(\mathbf{V}_\mathrm{RF}))= \mathrm{tr}(\mathbf{P}^{-1}\mathrm{d}(\mathbf{P})\mathbf{P}^{-1}).
\end{equation}
It can be further derived that
\begin{equation}\label{eqn:diff-P}
\begin{aligned}
\mathrm{d}(\mathbf{P}) = &\frac{1}{{\sigma^2w}}\mathbf{H}_\mathrm{1}^H\mathbf{V}_\mathrm{RF}(\mathrm{d}\left( {(\mathbf{V}_\mathrm{RF}^H{\mathbf{V}_\mathrm{RF}})^{ - 1}}\right)\mathbf{V}_\mathrm{RF}^H\\&+{(\mathbf{V}_\mathrm{RF}^H{\mathbf{V}_\mathrm{RF}})^{ - 1}}\mathrm{d}(\mathbf{V}_\mathrm{RF}^H))\mathbf{H}_\mathrm{1},
\end{aligned}
\end{equation}
where
\begin{equation}\label{eqn:diff-VRF^H-VRF}
\mathrm{d}\left({(\mathbf{V}_\mathrm{RF}^H{\mathbf{V}_\mathrm{RF}})^{ - 1}}\right)=-(\mathbf{V}_\mathrm{RF}^H{\mathbf{V}_\mathrm{RF}})^{ - 1}\mathrm{d}(\mathbf{V}_\mathrm{RF}^H)\mathbf{V}_\mathrm{RF}(\mathbf{V}_\mathrm{RF}^H{\mathbf{V}_\mathrm{RF}})^{ - 1}.
\end{equation}
By substituting (\ref{eqn:diff-VRF^H-VRF}) and (\ref{eqn:diff-P}) into (\ref{eqn:diff-J-derive-1}) and using again $\mathrm{tr}(\mathbf{AB})=\mathrm{tr}(\mathbf{BA})$, we have
\begin{equation}\label{eqn:diff-J-derive-2}
\begin{aligned}
\mathrm{d}(J(\mathbf{V}_\mathrm{RF}))= &\frac{1}{{\sigma^2w}} \mathrm{tr}((\mathbf{V}_\mathrm{RF} {(\mathbf{V}_\mathrm{RF}^H{\mathbf{V}_\mathrm{RF}})^{-1}} \mathbf{V}_\mathrm{RF}^H -\mathbf{I}_{N_\mathrm{t}})  \mathbf{H}_\mathrm{1}\\&\times\mathbf{P}^{-2} \mathbf{H}_\mathrm{1}^H \mathbf{V}_\mathrm{RF} {(\mathbf{V}_\mathrm{RF}^H{\mathbf{V}_\mathrm{RF}})^{-1}}\mathrm{d}(\mathbf{V}^H_\mathrm{RF})).
\end{aligned}
\end{equation}
By comparing (\ref{eqn:diff-J-derive-2}) with (\ref{eqn:diff-J-definition}), the proof is completed. \hfill $\square$

{ With the derived Euclidean conjugate gradient, the manifold optimization can be applied to solve the problem with the constant modulus constraints \cite{MO2009}. The overall MO-HBF algorithm is summarized in Algorithm 1, where the iteration index $i$ is denoted in the subscript of $\mathbf{V}_{\mathrm{RF},\,i}$. In particular, the detailed operation in the 4th step is given as follows.  First, project the Euclidean gradient onto the tangent space to obtain the Riemannian gradient. Second,  search a point in the tangent space along the Riemannian gradient and use the Armijo-Goldstein condition to determine the step size. Finally, retract the searched point back to the manifold.} %More detailed description of the MO method can be found in \cite{ZhangJun2016,MO2009}. The convergence of the MO-HBF algorithm will be proved in Section \ref{sec:convergence}.}

%\resizebox{0.9\textwidth}{!}{%
%	\begin{minipage}{1\textwidth}
\begin{algorithm}[t]
\caption{The MO-HBF Algorithm}
\hspace*{0.02in} {\bf Input:}
$\mathbf{H}_\mathrm{1}$, $\sigma^2$, $w$
\hspace*{0.02in} {\bf Output:}
$\mathbf{V}_\mathrm{RF}$, ${{\mathbf{V}}_\mathrm{U}}$, $\beta$
\begin{algorithmic}[1]
\State Initialize $\mathbf{V}_{\mathrm{RF},\,0} $ randomly and set $i=0$;
\State \textbf{repeat}	
\State Compute $\nabla J(\mathbf{V}_{\mathrm{RF},\, i})$ according to (\ref{eqn:gradient-J});
\State Use the manifold optimization method to compute $\mathbf{V}_{\mathrm{RF},(i+1)}$;
\State $i\leftarrow i+1$;
\State \textbf{Until} a stopping condition is satisfied;
\State Compute $\beta$ and ${{\mathbf{V}}_\mathrm{U}}$   according to (\ref{eqn:beta}) and (\ref{eqn:Lag-VD-result}).
\end{algorithmic}\label{alg:manifold}
\end{algorithm}
%\vspace{0.5cm}
%\end{minipage}%
%}

\subsubsection{Analog Precoder Design Based on the GEVD Method}\label{subsubsec:GEVD}
The above algorithm for optimizing the analog precoding matrix $\mathbf{V}_{\mathrm{RF}}$ in (\ref{pro:opt-V-RF}) is essentially a gradient based algorithm, where the computational complexity is proportional to the number of iterations and is related to the form of the objective function and the stop condition. In this part, we propose a low-complexity algorithm based on GEVD. According to \cite{YuWei2016}, for large-scale MIMO systems, it can be approximated that $\mathbf{V}_\mathrm{RF}^H\mathbf{V}_\mathrm{RF} \approx N_\mathrm{t}\mathbf{I}_{N_\mathrm{RF}}$ based on the fact that the optimized analog beamforming vectors for different streams are likely orthogonal to each other. With this approximation, (\ref{eqn:MSE-as-fun-VRF}) can be simplified as
\begin{equation}\label{eqn:MSE-orthogonal-approx}
J(\mathbf{V}_\mathrm{RF}) \approx\mathrm{tr}\left(\Big({\mathbf{I}_{N_{\mathrm{s}}} + \frac{1}{{\sigma^2wN_\mathrm{t}}}\mathbf{H}_\mathrm{1}^H\mathbf{V}_\mathrm{RF}{}\mathbf{V}_\mathrm{RF}^H \mathbf{H}_\mathrm{1}\Big)^{ - 1}}\right).
\end{equation}
With this simplified form, it can be shown that the analog precoding matrix $\mathbf{V}_\mathrm{RF}$ can be optimized column-by-column. Specifically, define $\overline{\mathbf{V}}_m$ as the remaining sub-matrix of $\mathbf{V}_\mathrm{RF}$ after removing the $m$th column $\mathbf{v}_m$. Further define $\mathbf{A}_{m} \triangleq  \mathbf{I}_{{N}_{\mathrm{s}}}+\frac{1}{{\sigma^2wN_\mathrm{t}}}\mathbf{H}_\mathrm{1}^H \overline{\mathbf{V}}_m\overline{\mathbf{V}}_m^H\mathbf{H}_\mathrm{1}$. Then, using the fact that $(\mathbf{A}+\mathbf{B})^{-1}=\mathbf{A}^{-1}-\frac{\mathbf{A}^{-1}\mathbf{B}\mathbf{A}^{-1}}{1+\mathrm{tr} (\mathbf{A}^{-1}\mathbf{B})}$ for a full-rank matrix $\mathbf{A}$ and a rank-one matrix $\mathbf{B}$, the MSE expression in (\ref{eqn:MSE-orthogonal-approx}) can be written as
\begin{align}
J(\mathbf{V}_\mathrm{RF}) &\approx \mathrm{tr}(\mathbf{A}^{-1}_{m})-\frac{\mathrm{tr}\left( \frac{1}{{\sigma^2wN_\mathrm{t}}}\mathbf{A}^{-1}_{m}\mathbf{H}_\mathrm{1}^H\mathbf{v}_m {}\mathbf{v}_m^{H}\mathbf{H}_\mathrm{1}\mathbf{A}^{-1}_{m}\right)}{1+\mathrm{tr}\left( \frac{1}{{\sigma^2wN_\mathrm{t}}}\mathbf{A}^{-1}_{m}\mathbf{H}_\mathrm{1}^H\mathbf{v}_m \mathbf{v}_m^{H}\mathbf{H}_\mathrm{1}\right)} \nonumber\\
&=\mathrm{tr}(\mathbf{A}^{-1}_{m})-\frac{\mathbf{v}_m^{H} \mathbf{U}_{{m}} \mathbf{v}_m}{\mathbf{v}_{m}^{H}\mathbf{W}_{m}\mathbf{v}_m}, \label{eqn:GEVD-each-step}
\end{align}

where $\mathbf{U}_{m} \triangleq\frac{1}{{\sigma^2wN_\mathrm{t}}}\mathbf{H}_\mathrm{1}\mathbf{A}^{-2}_{m} \mathbf{H}_\mathrm{1}^H$ and $\mathbf{W}_{m}\triangleq\frac{1}{N_\mathrm{t}}\mathbf{I}_{N_\mathrm{t}}+ \frac{1}{\sigma^2wN_\mathrm{t}}\mathbf{H}_\mathrm{1}\mathbf{A}^{-1}_{m}\mathbf{H}_\mathrm{1}^H$ are both  Hermitian  matrices. It is seen from (\ref{eqn:GEVD-each-step}) that the MSE expression is separated into two terms which are related to $\overline{\mathbf{V}}_{m}$ and $\mathbf{v}_m$, respectively. By fixing $\overline{\mathbf{V}}_m$, $J(\mathbf{V}_\mathrm{RF})$ becomes  a function on $\mathbf{v}_m$ in the second term in (\ref{eqn:GEVD-each-step}). As both $\mathbf{U}_{{m}}$ and $\mathbf{W}_{{m}}$ are Hermitian and $\mathbf{W}_{{m}}$ is  positive definite, according to \cite{Zhangxianda}, the optimal $\mathbf{v}_m$ in the sense of maximizing the last term in (\ref{eqn:GEVD-each-step}) or minimizing the whole term in (\ref{eqn:GEVD-each-step}) is the eigenvector associated with the maximum generalized eigenvalue between $\mathbf{U}_{{m}}$ and $\mathbf{W}_{{m}}$, which can be obtained via the GEVD operation. To further take the constant modulus constraint into account, a simple but effective way is to only extract the phase of each element in the generalized eigenvector. By applying the above GEVD and phase extraction operations for each column ${\mathbf{v}}_m$ and repeating them for the whole matrix ${\mathbf{V}}_\mathrm{RF}$ until the stop condition is satisfied, we finally get the optimized analog precoding matrix. The overall GEVD-HBF algorithm is summarized in Algorithm \ref{alg:GEVD}.

%\resizebox{0.9\textwidth}{!}{%
%	\begin{minipage}{1\textwidth}
\begin{algorithm}[t]
\caption{The GEVD-HBF Algorithm}
\hspace*{0.02in} {\bf Input:}
$\mathbf{H}_\mathrm{1}$, $\sigma^2$, $w$
\hspace*{0.02in} {\bf Output:}
 $\mathbf{V}_\mathrm{RF}$, ${{\mathbf{V}}_\mathrm{U}}$, $\beta$
\begin{algorithmic}[1]
\State Initialize $\mathbf{V}_\mathrm{RF,\,0} $ randomly and set $i$ = 0;
\For{$1\leq m\leq N_\mathrm{RF}$}
\State Compute $\mathbf{A}_{m}$, $\mathbf{U}_m$, $\mathbf{W}_m$ defined in Section \ref{subsubsec:GEVD};
\State Compute the maximum generalized eigenvector $\mathbf{z}$ of $\mathbf{U}_m$ and $\mathbf{W}_m$;
\State Set $\mathbf{v}_{m,\, i}=\mathrm{exp}\{\mathrm{j}\angle(\mathbf{z})\}$, i.e., extract the phase of each element of $\mathbf{z}$;
\EndFor
\State  $i\leftarrow i+1$;
\State Compute $\beta$ and ${{\mathbf{V}}_\mathrm{U}}$  according to (\ref{eqn:beta}) and (\ref{eqn:Lag-VD-result}) .
\end{algorithmic}\label{alg:GEVD}
\end{algorithm}
%\end{minipage}%
%\begin{minipage}{1\textwidth}
%\begin{algorithm}[H]
%	\caption{The MO-HBF Algorithm}
%	\hspace*{0.02in} {\bf Input:}
%	$\mathbf{H}_\mathrm{1}$, $\sigma^2$, $w$\\
%	\hspace*{0.02in} {\bf Output:}
%	$\mathbf{V}_\mathrm{RF}$, ${{\mathbf{V}}_\mathrm{U}}$, $\beta$
%	\begin{algorithmic}[1]
%		\State Initialize $\mathbf{V}_{\mathrm{RF},\,0} $ randomly and set $i=0$;
%		\State \textbf{repeat}	
%		\State Compute $\nabla J(\mathbf{V}_{\mathrm{RF},\, i})$ according to (\ref{eqn:gradient-J});
%		\State Use the manifold optimization method to compute $\mathbf{V}_{\mathrm{RF},(i+1)}$;
%		\State $i\leftarrow i+1$;
%		\State \textbf{Until} a stopping condition is satisfied;
%		\State Compute $\beta$ and ${{\mathbf{V}}_\mathrm{U}}$   according to (\ref{eqn:beta}) and (\ref{eqn:Lag-VD-result}).
%	\end{algorithmic}\label{alg:manifold}
%\end{algorithm}
%\end{minipage}%

\subsection{Hybrid Receive Combiner Design}\label{subsec:combining}
By fixing the updated precoding matrices ${\mathbf{V}}_{\mathrm{U}}$ and $\mathbf{V}_{\mathrm{RF}}$ along with the scaling factor $\beta$, the optimizing problem in (\ref{pro:opt-total}) can be reduced to the following one for the hybrid receive combiner
\begin{equation}\label{pro:opt-combining}
\begin{array}{cl}
\displaystyle{\minimize_{\mathbf{W}_\mathrm{RF},{\mathbf{W}}_\mathrm{B}}} & \mathrm{tr} 	({\mathbf{W}}^H\mathbf{H}_\mathrm{2}{\mathbf{H}}_\mathrm{2}^H{\mathbf{W}} -{\mathbf{W}}^H \mathbf{H}_\mathrm{2} - \mathbf{H}_\mathrm{2}^H{\mathbf{W}} \\ &\quad+ {\sigma^2}\beta^{-2}{\mathbf{W}}^H{\mathbf{W}}+\mathbf{I}_{N_\mathrm{s}} )\\ \\
\mathrm{subject \; to} & \quad|[\mathbf{W}_{\mathrm{RF}}]_{ml}|=1, \forall m,l,
\end{array}
\end{equation}
where $\mathbf{H}_\mathrm{2} \triangleq \mathbf{H}\mathbf{V}_\mathrm{RF}{{\mathbf{V}}_\mathrm{U}}$ and ${\mathbf{W}} \triangleq {\mathbf{W}}_\mathrm{RF}{\mathbf{W}}_\mathrm{B}$. Similarly, by differentiating the objective function of (\ref{pro:opt-combining}) with respect to ${\mathbf{W}}_\mathrm{B}$ and setting the result to zero, we have the optimal ${\mathbf{W}}_\mathrm{B}$ as follows
\begin{equation}\label{eqn:WD}
{\mathbf{W}}_\mathrm{B} = (\mathbf{W}_\mathrm{RF}^H\mathbf{H}_\mathrm{2}{\mathbf{H}}_\mathrm{2}^H\mathbf{W}_\mathrm{RF}+ {\sigma^2}\beta^{-2}\mathbf{W}_\mathrm{RF}^H\mathbf{W}_\mathrm{RF})^{-1}\mathbf{W}_\mathrm{RF}^H \mathbf{H}_\mathrm{2}.
\end{equation}
Substituting (\ref{eqn:WD}) back into the problem in (\ref{pro:opt-combining}), we have

\begin{equation} \label{pro:opt-W-RF}
\begin{array}{cl}
\displaystyle{\minimize_{\mathbf{W}_\mathrm{RF}}} &\quad I(\mathbf{W}_\mathrm{RF}) \triangleq\mathrm{tr}((\mathbf{I}_{{N}_{\mathrm{s}}} + \sigma^{-2}\beta^{2} \mathbf{H}_\mathrm{2}^H\mathbf{W}_\mathrm{RF}\\ &\quad\quad\quad\quad\quad\times{(\mathbf{W}_\mathrm{RF}^H {\mathbf{W}_\mathrm{RF}})^{ - 1}}\mathbf{W}_\mathrm{RF}^H\mathbf{H}_\mathrm{2})^{ - 1})\\
\mathrm{subject \; to} &\quad|[\mathbf{W}_{\mathrm{RF}}]_{ml}|=1, \forall m,l.
\end{array}
\end{equation}
Comparing (\ref{eqn:WD}) with (\ref{eqn:Lag-VD-result}) and (\ref{pro:opt-W-RF}) with (\ref{pro:opt-V-RF}), it can be seen that they have almost the same form respectively. Thus, the MO-HBF and GEVD-HBF algorithms, which were introduced in Section~\ref{subsec:precoding}, can be directly applied to optimize the hybrid combiner.

% Specifically, using $\mathbf{H}_\mathrm{2}$  and $\beta$ as the input `$\mathbf{H}_\mathrm{2}$' and `$w$' respectively  of Algorithm 1 and Algorithm 2,  then the ouputs `$\beta$',`$\mathbf{V}_\mathrm{RF}$', '${{\mathbf{V}}_\mathrm{U}}$' are exactly the value of $w,\mathbf{W}_\mathrm{RF},{\mathbf{W}}_\mathrm{U}$.  Due to the duality of the optimization at the transmitter and the receiver, the proposed precoding algorithm can effectively form the combiners. A pair of ${\mathbf{W}}_\mathrm{U}$ and $\mathbf{W}_\mathrm{RF}$ can be obtained in this way and  $w$  can be used to form the hybrid precoders at the transmitter side in the next itertion.

\subsection{Alternating Minimization for  Hybrid Beamforming}\label{subsec:alternating}
\subsubsection{Alternating Optimization}
A joint hybrid precoding and combining design based on the MMSE criterion can be developed by iteratively and alternatively using the hybrid precoding design in Section \ref{subsec:precoding} and the hybrid combining design in Section \ref{subsec:combining}. Specifically, during the $n$th iteration, first for the optimization of the hybrid precoder, by updating the problem in (\ref{pro:opt-precoding}) with the optimized combiners ${\mathbf{W}}_\mathrm{RF}^{(n-1)}$ and ${\mathbf{W}}_\mathrm{B}^{(n-1)}$ in the $(n-1)$th iteration, the hybrid precoding matrices $\mathbf{V}_\mathrm{RF}^{(n)}$, ${{\mathbf{V}}_\mathrm{U}}^{(n)}$ and the scaling factor $\beta^{(n)}$ are optimized via the MO-HBF  or the GEVD-HBF algorithm. Similarly, with the new hybrid precoder, the hybrid combining optimization problem in (\ref{pro:opt-combining}) is then updated and solved via the same algorithm. This alternating optimization is repeated until a stop condition is satisfied.

\subsubsection{Stopping Condition}\label{stopping condition}
{To distinguish the iteration of the alternating minimization between the transmitter and receiver beamforming optimization and the iteration in the optimization of the analog beamformer (i.e., Algorithms 1 and 2), we refer to the former as the outer iteration and the latter as the inner iteration. The stopping condition of these two iterations can be set as either the number of iterations exceeding a specified value $\alpha$, or the relative difference between the MSE values of two consecutive iterations becoming smaller than a specified value $\delta$. For example, considering a typical system configuration in Section VII, according to the observation in simulations, good performance can be achieved when we set $\delta=10^{-5}$ for both the inner and outer iterations for the alternating MO-HBF algorithm and set $\alpha=1$ for the inner iteration and $\delta=10^{-5}$ for the outer iteration for the alternating GEVD-HBF algorithm.}

\subsubsection{Beamforming Initialization}\label{Initialization}
It is worth noting that the number of iterations for the alternating optimization highly depends on the initialization of the beamformers. One simple idea is to randomly generate a hybrid combiner (precoder) for the optimization of the precoder (combiner) at the beginning of the alternating minimization. This method does not require extra information, but may need a lot of iterations to converge to a local optimal point with certain performance loss. As the concatenation of the hybrid beamformer will gradually approach the full-digital one during the iterations, we propose to take a full-digital beamformer as the initialization for better convergence. Specifically, the optimal full-digital precoder based on the MMSE criterion proposed in \cite{Smapth:2001} can be used here. Assuming without loss of generality that the alternating optimization starts from the transmit beamforming problem in (\ref{pro:opt-precoding}), we assume that there is a virtual full-digital combiner at the receiver in the initialization step. Namely, we initialize the concatenation of the hybrid combiner, ${\mathbf{W}}^{(0)}$, as the optimal full-digital combiner in \cite{Smapth:2001} and substitute it into (\ref{pro:opt-precoding}) for the precoder optimization in the first iteration. We refer to the proposed initialization method as the virtual full-digital beamformer (VFD) method. { It is worth noting that as the VFD initialization method assumes a virtual full-digital beamformer at one side, which generally cannot be directly implemented using the HBF structure, at least one outer iteration is needed to obtain the hybrid beamformers for both sides.} As the VFD initialization does not require the alternating optimization to obtain the full-digital beamformer, its additional complexity is much lower compared with that of the main HBF algorithms. Simulations in Section \ref{sec:simulation} will show that with the VFD method, the convergence speed improves significantly with even some MSE performance improvement, in comparison with random initialization.

%\begin{algorithm}[t]
%\caption{ Alternative algorithm for hybrid beamforming.}
%\hspace*{0.02in} {\bf Input:}
% $\sigma^2$,$\mathbf{V}_\mathrm{opt}$\\
%\hspace*{0.02in} {\bf Output:}
%$\mathbf{V}_\mathrm{RF}$, ${{\mathbf{V}}_\mathrm{U}}$,$\mathbf{W}_\mathrm{RF}$, ${\mathbf{W}}_\mathrm{U}$
%\begin{algorithmic}[1]
%\State Initialize ${\mathbf{W}}^{\mathrm{(0)}} =\mathbf{W}_\mathrm{opt}$ and $n$ = 1;
%\State \textbf{repeat}
%\State Compute $\mathbf{H}_{\mathrm{1}}^{(n)} = \mathbf{H}^H{\mathbf{W}}^{(n-1)}$
%\State Obtain $\mathbf{V}_{\mathrm{RF}}^{(n)}$, ${\mathbf{V}}_{\mathrm{U}}^{(n)}$ and $\beta^{\mathrm{(n)}}$.
%\State Compute $\mathbf{H}_{\mathrm{2}}^{(n)} = \mathbf{H}\mathbf{V}^{(n)}$.
%\State Obtain $\mathbf{W}_{\mathrm{RF}}^{(n)}$ and ${\mathbf{W}}_{\mathrm{U}}^{(n)}$.
%\State  $n\leftarrow n+1$
%\State \textbf{Until} a stopping criterion triggers.
%\end{algorithmic}\label{alg:overall}
%\end{algorithm}

\section{Hybrid BeamForming Design for Broadband MmWave Systems}\label{sec:design-broadband}
\begin{figure*}[!t]
\centering
\includegraphics[width=6.6in]{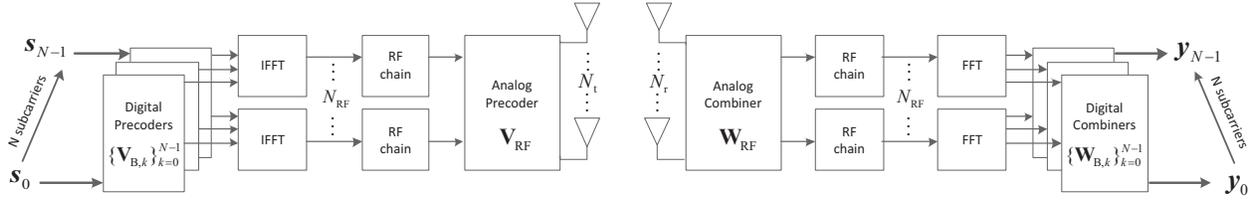}%or 8.5cm
\caption{Diagram of a broadband mmWave MIMO system with HBF.}
\label{fig:sys-broadband}
\vspace{-0.5cm}
\end{figure*}
Due to the large available bandwidth of mmWave systems, frequency selective fading will be encountered. Therefore, in this section we generalize the previous hybrid beamformer design to broadband mmWave systems. %After introducing the system model and formulating the optimization problem, we show the similarity and differences between the broadband and narrowband scenarios.
In particular, we point out that similar to the narrowband scenario, the optimization of precoder and combiner in the broadband scenario can also be unified  and solved through the same procedure. Thus, we focus on the hybrid transmit precoder design in the broadband scenario and propose three algorithms.

\subsection{System Model in the Broadband Scenario}\label{subsec:system-model-broadband}
To overcome the channel frequency selectivity, we assume that the orthogonal frequency division multiplexing (OFDM) technology is applied so that the channel fading on each subcarrier can be regarded as being flat. To facilitate the following system design, the broadband mmWave MIMO channel model with half-wavelength spaced ULAs at both the transmitter and the receiver in \cite{YuWei2017} is adopted here, where the channel matrix at the $k$th subcarrier, for $k=0,\ldots,N-1$ with $N$ being the total number of subcarriers, is given by
\begin{equation}\label{eqn:Hmodel-broad}
\mathbf{H}_k = \sqrt {\frac{{{N_\mathrm{t}}{N_\mathrm{r}}}}{{{N_\mathrm{C}}{N_{\mathrm{R}}}}}} \sum\limits_{i = 1}^{{N_\mathrm{C}}} {\sum\limits_{j = 1}^{{N_{\mathrm{R}}}} {{\alpha_{ij}}} } {\mathbf{a}_\mathrm{r}}(\theta _{ij}^\mathrm{r}){\mathbf{a}_\mathrm{t}}{(\theta_{ij}^\mathrm{t})}^He^{ - \mathrm{j} \frac{2\pi}{N}(i-1)k},
\end{equation}
where the other parameters are defined in the same way as that in (\ref{eqn:Hmodel-narrow}). It is worth noting that although the geometry-based spatial channel model is applied in simulations, all proposed HBF algorithms are compatible for other general models.

%\begin{figure*}[!t]
%\centering
%\includegraphics[width=7in]{broadbandsystem.eps}%or 8.5cm
%\caption{Diagram of an broadband mmWave MIMO system with hybrid beamforming with power normalizing factor }
%\label{broadbandsystem}
%\end{figure*}
As shown in Fig. \ref{fig:sys-broadband}, the processed signal vector at the $k$th subcarrier after the hybrid receive combining can be expressed as
\begin{equation}\label{eqn:y-broadband}
\mathbf{y}_k=\mathbf{W}_{\mathrm{B},k}^H\mathbf{W}_\mathrm{RF}^H\mathbf{H}_k\mathbf{V}_\mathrm{RF} \mathbf{V}_{\mathrm{B},k}\mathbf{s}_k+ \mathbf{W}_{\mathrm{B},k}^H\mathbf{W}_\mathrm{RF}^H\mathbf{u}_k,
\end{equation}
where $\mathbf{s}_k$ and $\mathbf{u}_k$ denote the transmitted symbol vector and the additive noise vector at the $k$th subcarrier, respectively, $\mathbf{V}_{\mathrm{B},k}$ and $\mathbf{W}_{\mathrm{B},k}$ denote the digital precoder and combiner at the $k$th subcarrier, respectively, and $\mathbf{V}_\mathrm{RF}$ and $\mathbf{W}_\mathrm{RF}$ denote the analog precoder and combiner, respectively. It is worth noting that in the broadband scenario, the digital beamformers $\mathbf{V}_{\mathrm{B},k}$ and $\mathbf{W}_{\mathrm{B},k}$ must be optimized  for different subcarriers while the analog precoder or combiner is invariant for all subcarriers due to the post-IFFT or pre-FFT processing. Similar to \cite{ZhangJun2016,YuWei2017}, we assume equal power allocation among subcarriers at the transmitter, namely $\|\mathbf{V}_\mathrm{RF}\mathbf{V}_{\mathrm{B},k}\|_F^2\leq 1, \mathrm{for} \; k=0,\ldots,N-1$.  To deal with  the new difficulty in the HBF design for broadband mmWave MIMO systems, we take the sum-MSE of all the subcarriers and all the streams as the objective function. That is, define
$\mathrm{MSE}=\sum_{k = 0}^{N - 1}\mathrm{MSE}_k$, where $\mathrm{MSE}_k$ denotes the modified MSE  on the $k$th subcarrier and is given by
\begin{equation}
\begin{aligned}
\mathrm{MSE}_k&=E(\|\mathbf{s}_k-\beta^{-1}_k\mathbf{y}_k\|^2)\\
&=\mathrm{tr} 	(\beta_k^{-2}\mathbf{W}_k^H\mathbf{H}_k\mathbf{V}_k\mathbf{V}_k^H\mathbf{H}_k^H\mathbf{W}_k  -\beta_k^{-1}\mathbf{W}_k^H\mathbf{H}_k\mathbf{V}_k  \\&\quad\quad-\beta_k^{-1}\mathbf{V}_k^H\mathbf{H}_k^H\mathbf{W}_k  + {\sigma^2}\beta_k^{-2}\mathbf{W}_k^H\mathbf{W}_k + \mathbf{I}_{N_\mathrm{s}}),
\end{aligned}
\end{equation}
where ${\mathbf{W}_k} \triangleq \mathbf{W}_{\mathrm{RF}} \mathbf{W}_{\mathrm{B},k}$, ${\mathbf{V}_k}\triangleq \mathbf{V}_{\mathrm{RF}} \mathbf{V}_{\mathrm{B},k}$, and $\beta_k$ is a scaling factor for the $k$th subcarrier as similar to that in the narrowband scenario. Then, the optimization problem in the broadband scenario can be formulated as
\begin{equation}\label{pro:opt-broadband}
\begin{array}{cl}
\displaystyle{\minimize_{\mathbf{V}_{\mathrm{B},k},\mathbf{V}_{\mathrm{RF}},\mathbf{W}_{\mathrm{U},k}, \mathbf{W}_{\mathrm{RF}},\beta_k}} & \sum_{k = 0}^{N - 1}\mathrm{MSE}_k\\
\mathrm{subject \; to} & \|\mathbf{V}_k\|_F^2\leq 1,\quad \\&|[\mathbf{V}_{\mathrm{RF}}]_{ij}|=1, \forall i,j, \quad \\& |[\mathbf{W}_{\mathrm{RF}}]_{ml}|=1, \forall m,l.
\end{array}
\end{equation}
By comparing the problem in (\ref{pro:opt-broadband}) with that in (\ref{pro:opt-total}) for the narrowband scenario, it can be found that they have almost the same form except that the digital beamformers need to be optimized for different subcarriers in (\ref{pro:opt-broadband}). Thus, the alternating minimization principle is also applicable here. In particular, it can be shown that in the broadband scenario the  two sub-problems associated with (\ref{pro:opt-broadband}) for the optimization of precoding and combining can also be solved through the same procedure. Therefore, in the following we focus on the hybrid transmit precoder design.

\subsection{Broadband Hybrid Transmitter Design}\label{subsec:precoding-broadband}
Analogous to that in Section \ref{subsec:precoding}, the original precoder $\mathbf{V}_{\mathrm{B},k}$ is also separated as $\mathbf{V}_{\mathrm{B},k}=\beta_k\mathbf{V}_{\mathrm{U},k}$, where $\mathbf{V}_{\mathrm{U},k}$ is an unnormalized  precoder for the $k$th subcarrier. It can also be proved by contradiction that the optimal $\beta_k$ is given by $  \left(\mathrm{tr}(\mathbf{V}_\mathrm{RF}{\mathbf{V}}_{\mathrm{U},k} {\mathbf{V}}_{\mathrm{U},k}^H\mathbf{V}_\mathrm{RF}^H)\right)^{-\frac{1}{2}}$. Then, by fixing the hybrid receive combiner and based on the KKT conditions, the optimal ${\mathbf{V}}_{\mathrm{U},k}$ can be derived as a function of $\mathbf{V}_\mathrm{RF}$. That is,
\begin{equation}\label{eqn:VD-k-broad}
{\mathbf{V}}_{\mathrm{U},k} =(\mathbf{V}_\mathrm{RF}^H\mathbf{H}_{\mathrm{1},k}\mathbf{H}_{\mathrm{1},k}^H\mathbf{V}_\mathrm{RF}+ {\sigma^2}w_k\mathbf{V}_\mathrm{RF}^H\mathbf{V}_\mathrm{RF})^{-1}\mathbf{V}_\mathrm{RF}^H \mathbf{H}_{\mathrm{1},k},
\end{equation}
where $\mathbf{H}_{\mathrm{1},k}\triangleq\mathbf{H}^H_k {\mathbf{W}_k}$ and $w_k \triangleq \mathrm{tr}({\mathbf{W}_k}{\mathbf{W}_k}^H)$. Now the original problem in (\ref{pro:opt-broadband}) is reduced to the one for $\mathbf{V}_\mathrm{RF}$ as follows
\begin{equation}\label{pro:opt-broad-J}
\begin{array}{cl}
\displaystyle{\minimize_{\mathbf{V}_\mathrm{RF}}} & J(\mathbf{V}_\mathrm{RF})=\sum\limits_{k=0}^{N-1} J_k(\mathbf{V}_\mathrm{RF}) \\
\mathrm{subject \; to} & |(\mathbf{V}_{\mathrm{RF}}) _{ij}|^2=1, \forall i,j,
\end{array}
\end{equation}
where  
\begin{equation}
J_k(\mathbf{V}_\mathrm{RF}) \triangleq \mathrm{tr}({(\mathbf{I}_{{N}_{\mathrm{s}}} + \frac{1}{{\sigma^2w_k}}\mathbf{H}_{\mathrm{1},k}^H\mathbf{V}_\mathrm{RF} {(\mathbf{V}_\mathrm{RF}^H{\mathbf{V}_\mathrm{RF}})^{ - 1}}\mathbf{V}_\mathrm{RF}^H\mathbf{H}_{\mathrm{1},k})^{ - 1}})
\end{equation}
%\begin{equation}\label{eqn:def-J-k}
% J_k(\mathbf{V}_\mathrm{RF}) \triangleq \mathrm{tr}\Big({\big(\mathbf{I}_{{N}_{\mathrm{s}}} + \frac{1}{{\sigma^2w_k}}\mathbf{H}_{\mathrm{1},k}^H\mathbf{V}_\mathrm{RF} {(\mathbf{V}_\mathrm{RF}^H{\mathbf{V}_\mathrm{RF}})^{ - 1}}\mathbf{V}_\mathrm{RF}^H\mathbf{H}_{\mathrm{1},k}\big)^{ - 1}}\Big).
%\end{equation}

\subsubsection{Analog Precoding Based on the MO Method}
Although the objective function for analog precoding optimization in the broadband scenario is more complicated than that in the narrowband scenario, the MO method can still be applied here. Using some differentiation rules for complex-valued matrices, the conjugate gradient of the function $J(\mathbf{V}_\mathrm{RF})$ with respect to $\mathbf{V}_\mathrm{RF}$ can be expressed as
\begin{equation}\label{eqn:gW-broadband}
\nabla J(\mathbf{V}_\mathrm{RF}) =  \sum\limits_{k = 0}^{N - 1}\nabla J_k(\mathbf{V}_\mathrm{RF}),
\end{equation}
where $\nabla J_k(\mathbf{V}_\mathrm{RF})$ can be derived as follows according to Lemma \ref{lemma:MO}
\begin{equation}\label{eqn:gradient-J2}
\begin{aligned}
\nabla J_k(\mathbf{V}_\mathrm{RF})= &\frac{1}{{\sigma^2w_k}} \left(\mathbf{V}_\mathrm{RF} {\left(\mathbf{V}_\mathrm{RF}^H{\mathbf{V}_\mathrm{RF}}\right)^{-1}} \mathbf{V}_\mathrm{RF}^H -\mathbf{I}_{N_\mathrm{RF}}\right)\\&\times  \mathbf{H}_{\mathrm{1},k}\mathbf{P}_k^{-2} \mathbf{H}_{\mathrm{1},k}^H \mathbf{V}_\mathrm{RF} {\left(\mathbf{V}_\mathrm{RF}^H{\mathbf{V}_\mathrm{RF}}\right)^{-1}},
\end{aligned}
\end{equation}
with $\mathbf{P}_k \triangleq \mathbf{I}_{\mathrm{N}_{\mathrm{s}}} + \frac{1}{{\sigma^2w_k}}\mathbf{H}_{\mathrm{1},k}^H\mathbf{V}_\mathrm{RF} {(\mathbf{V}_\mathrm{RF}^H{\mathbf{V}_\mathrm{RF}})^{ - 1}}\mathbf{V}_\mathrm{RF}^H\mathbf{H}_{\mathrm{1},k}$ defined for notational brevity. Thus, the Riemannian gradient can be computed by projecting the above Euclidean gradient onto the tangent space of the Riemannian manifold \cite{ZhangJun2016}. According to the property of the gradient descent method, with a proper selection of the step size, $\mathbf{V}_\mathrm{RF}$ is guaranteed to converge to a feasible local optimal solution via MO.

%\begin{algorithm}[t]
%\caption{ Precoding algorithm for broad-band systems.}
%\hspace*{0.02in} {\bf Input:}
%$\mathbf{H}_{\mathrm{1},k}$, $\sigma^2$,$w_k$\\
%\hspace*{0.02in} {\bf Output:}
% $\mathbf{V}_\mathrm{RF}$, ${\mathbf{V}}_{\mathrm{U},k}$, $\beta_k$
%\begin{algorithmic}[1]
%\State Initialize $\mathbf{V}_{\mathrm{RF}} $ randomly.
%\State \textbf{repeat}
%\For{$0\leq k\leq N-1$}
%\State Calculate $\nabla J_k(\mathbf{V}_\mathrm{RF})$ according to (\ref{eqn:g(W)ij}),  (\ref{eqn:g(W)ij1}), (\ref{eqn:g(W)ij2}).
%\EndFor
%\State Calculate $ \nabla J(\mathbf{V}_\mathrm{RF}) = \sum\limits_{k = 0}^{N - 1}\nabla J_k(\mathbf{V}_\mathrm{RF})$,
%\State Using the manifold optimization in \cite{ZhangJun2016} to update  $\mathbf{V}_\mathrm{RF} $.
%\State \textbf{Until} a stopping criterion triggers.
%\State Set ${\mathbf{V}}_{\mathrm{U},k}=\beta_k^{-\frac{1}{2}}(\mathbf{V}_\mathrm{RF}^H\mathbf{H}_{\mathrm{1},k}\mathbf{H}_{\mathrm{1},k}^H\mathbf{V}_\mathrm{RF}+{\sigma^2}w_k\mathbf{V}_\mathrm{RF}^H\mathbf{V}_\mathrm{RF})^{-1}\mathbf{V}_\mathrm{RF}^H\mathbf{H}_{\mathrm{1},k}$  \\Set $\beta_k={\mathrm{tr}\left(\mathbf{V}_\mathrm{RF}^H{\mathbf{V}}_{\mathrm{U},k}^H{\mathbf{V}}_{\mathrm{U},k}\mathbf{V}_\mathrm{RF}\right)}$.
%\end{algorithmic}
%\end{algorithm}
\subsubsection{Analog Precoding Based on the EVD Method}\label{subsubsec:low-com-broadband}
%Although the MO-HBF algorithm can achieve good performance, the objective function in the broadband scenario is more complicated than that in the narrowband scenario, which leads to higher computational complexity. Therefore, low-complexity analog precoding algorithm needs to be investigated. Unfortunately, the GEVD-HBF algorithm in Section \ref{subsubsec:GEVD} does not work here. Since the objective function in the broadband scenario is the sum-MSE of all subcarriers, the variant channel matrices and digital beamformers at different subcarriers prevent us from rewriting the original problem in an GEVD-available formulation. Nevertheless, the approximation of $\mathbf{V}_\mathrm{RF}^H\mathbf{V}_\mathrm{RF}\approx {N_\mathrm{t}}\mathbf{I}_{N_\mathrm{RF}}$ can still be utilized here for developing another low-complexity algorithm.
Note that since the objective function in the broadband scenario is the sum-MSE of all the subcarriers, the variant channel matrices and digital beamformers at different subcarriers prevent us from rewriting the original problem in an GEVD-available formulation as that in Section \ref{subsubsec:GEVD}. Nevertheless, the approximation of $\mathbf{V}_\mathrm{RF}^H\mathbf{V}_\mathrm{RF}\approx {N_\mathrm{t}}\mathbf{I}_{N_\mathrm{RF}}$ can still be utilized here for developing other low-complexity algorithms. The basic idea is to ignore the constant modulus constraint in (\ref{pro:opt-broad-J}) temporarily and add a new constraint of $\mathbf{V}_\mathrm{RF}^H\mathbf{V}_\mathrm{RF}={N_\mathrm{t}} \mathbf{I}_{N_\mathrm{RF}}$. Then, we have the following new problem
\begin{equation}\label{Pro:opt-EVD-broadband}
\begin{array}{cl}
\displaystyle{\minimize_{\mathbf{V}_\mathrm{RF}}} &
J(\mathbf{V}_\mathrm{RF})=\sum_{k = 0}^{N - 1}
\mathrm{tr}((\mathbf{I}_{{N}_{\mathrm{s}}} +
\frac{1}{\sigma^2w_kN_\mathrm{t}} \\& \quad\quad\quad\quad\quad\times\mathbf{H}_{\mathrm{1},k}^H
\mathbf{V}_\mathrm{RF} \mathbf{V}_\mathrm{RF}^H
\mathbf{H}_{\mathrm{1},k})^{-1}) \\
\mathrm{subject \; to} &
\mathbf{V}_\mathrm{RF}^H\mathbf{V}_\mathrm{RF}={N_\mathrm{t}}\mathbf{I}_{N_\mathrm{RF}}.
\end{array}
\end{equation}
It turns out that the above optimization problem is still difficult to solve. Therefore, we
devote to derive its lower bound first with the help of the following lemma.

\begin{lemma}
A lower bound of the objective in (\ref{Pro:opt-EVD-broadband}) is given by
	\begin{equation}
	J(\mathbf{V}_\mathrm{RF})\ge\frac{N^2N_\mathrm{s}^2}{\sum_{k = 0}^{N-1}\mathrm{tr}(\mathbf{I}_{{N}_{\mathrm{s}}} + \frac{1}{{\sigma^2w_kN_\mathrm{t}}} \mathbf{H}_{\mathrm{1},k}^H\mathbf{V}_\mathrm{RF}{}\mathbf{V}_\mathrm{RF}^H\mathbf{H}_{\mathrm{1},k})}.
	\end{equation}
\end{lemma}
\textit{Proof}: For notational brevity, we define $\mathbf{Q}_k\triangleq\mathbf{I}_{{N}_{\mathrm{s}}} + \frac{1}{{\sigma^2w_kN_\mathrm{t}}} \mathbf{H}_{\mathrm{1},k}^H\mathbf{V}_\mathrm{RF}{}\mathbf{V}_\mathrm{RF}^H\mathbf{H}_{\mathrm{1},k}$ and have
\begin{equation} \label{eqn:EVD-JS1}
\begin{aligned}
J(\mathbf{V}_\mathrm{RF})&=\sum_{k = 0}^{N-1}\mathrm{tr}(\mathbf{Q}_k^{-1}) =  \sum_{k = 0}^{N-1}\sum_{i = 1}^{{N}_{\mathrm{s}}}\frac{1}{\lambda_{i,k}}\mathop  \ge \limits^{(a)}  \sum_{k = 0}^{N-1}\frac{{N}_{\mathrm{s}}^2}{\sum_{i = 1}^{{N}_{\mathrm{s}}}\lambda_{i,k}} \\&=\sum_{k = 0}^{N-1} \frac{{N}_{\mathrm{s}}^2}{\mathrm{tr}(\mathbf{Q}_k)}\mathop  \ge \limits^{(b)} \frac{N^2{N}_{\mathrm{s}}^2}{\sum_{k = 0}^{N-1}\mathrm{tr}(\mathbf{Q}_k)} ,
\end{aligned}
\end{equation}
where $\lambda_{i,k}$ denotes the $i$th eigenvalue of $\mathbf{Q}_k$ and is positive because $\mathbf{Q}_k$ is positive definite. The inequalities (a) and (b) in (\ref{eqn:EVD-JS1}) both come from the Jensen's inequality, with  equality of (a) satisfied if $\lambda_{1,k} = \lambda_{2,k} =\ldots=\lambda_{{N}_{\mathrm{s}},k}$ and equality of (b) satisfied if $\mathrm{tr}(\mathbf{Q}_0)=\mathrm{tr}(\mathbf{Q}_1)=\ldots=\mathrm{tr}(\mathbf{Q}_{N-1})$, respectively. Substituting the definition of $\mathbf{Q}_k$, the proof is completed. \hfill $\square$

Then, instead of the objective function in problem  (\ref{Pro:opt-EVD-broadband}), we devote to minimize its lower-bound, which is equivalent to maximizing  ${\sum_{k = 0}^{N-1}\mathrm{tr}(\mathbf{Q}_k)}$. After omitting the constant terms, the optimization problem can be rewritten as
\begin{equation}\label{pro:opt-low-broadband-approx}
\begin{array}{cl}
\displaystyle{\maximize_{\mathbf{V}_\mathrm{RF}}} & \mathrm{tr}\left(\mathbf{V}_\mathrm{RF}^H\left(\sum_{k = 0}^{N-1}\mathbf{H}_{\mathrm{1},k}\mathbf{H}_{\mathrm{1},k}^H\right)\mathbf{V}_\mathrm{RF}\right) \\
\mathrm{subject \; to} & \mathbf{V}_\mathrm{RF}^H\mathbf{V}_\mathrm{RF}={N_\mathrm{t}}\mathbf{I}_{N_\mathrm{RF}}.
\end{array}
\end{equation}
It can be proved that the optimal $\mathbf{V}_\mathrm{RF}$ is $\sqrt{N_\mathrm{t}}$ times the isometric matrix containing the $N_\mathrm{RF}$ eigenvectors associated with the largest $N_\mathrm{RF}$  eigenvalues of $\left(\sum_{k = 0}^{N - 1}\mathbf{H}_{\mathrm{1},k}\mathbf{H}_{\mathrm{1},k}^H\right)$ \cite{KFAN}, which can be obtained through EVD. To further make the constant modulus constraint satisfied, similar to that in Section \ref{subsec:precoding}, we just extract the phase of each element of the optimal $\mathbf{V}_\mathrm{RF}$. { The algorithm is referred to as the EVD-LB-HBF algorithm, where LB denotes the abbreviation for lower bound. In the following, we propose a better algorithm, where instead of minimizing a lower bound of (\ref{Pro:opt-EVD-broadband}) in EVD-LB-HBF an upper bound is derived for minimization.
%is not intuitive. Nevertheless, though the following lemma, we can obtain the upper bound of (\ref{Pro:opt-EVD-broadband}).

\begin{lemma}\label{Lemma:UB}
For an $a\times a$ positive definite and Hermitian matrix $\mathbf{A}$ and an arbitrary $a\times b$  $(a>b)$  para-unitary matrix $\mathbf{B}$, i.e., $\mathbf{B}^H\mathbf{B}=\mathbf{I}_n$, define the eigenvalues of $(\mathbf{B}^H\mathbf{A}\mathbf{B})^{-1}$ and $\mathbf{B}^H\mathbf{A}^{-1}\mathbf{B}$ in descending order as $\mu_1,...,\mu_{n}$ and $\lambda_1,...,\lambda_{n}$, respectively. Then we have $\mu_k \le \lambda_k, \forall k$.
\end{lemma}
\textit{Proof}: According to Courant-Fisher min-max theorem \cite{1990Matrix},
\begin{equation}
\lambda_k = \mathop {\max }\limits_\mathbb{U} \mathop {\min
}\limits_{\mathbf{x} \in \mathbb{U}}
\frac{\mathbf{x}^H\mathbf{B}^H\mathbf{A}^{-1}\mathbf{B}\mathbf{x}}{\mathbf{x}^H\mathbf{x}}
=\mathop {\max }\limits_\mathbb{U} \mathop {\min }\limits_{\mathbf{x} \in
\mathbb{F}}
\frac{\mathbf{x}^H\mathbf{A}^{-1}\mathbf{x}}{\mathbf{x}^H\mathbf{x}}, \nonumber
\end{equation}
where $\mathbf{x}$ is a non-zero vector, $\mathbb{U}$ denotes a $k$-dimension subspace of $\mathbb{C}^m$ and $\mathbb{F}$ is a new subspace after a linear transform of $\mathbf{B}$ to $\mathbb{U}$. Similarly, as $1{\rm{/}}\mu_k$ can be proved to be the $(b-k+1)$th largest eigenvalue of $\mathbf{B}^H\mathbf{A}\mathbf{B}$, we have
\begin{equation}
\frac{1}{\mu_k} = \mathop {\min }\limits_\mathbb{U} \mathop {\max
}\limits_{\mathbf{x} \in \mathbb{U}}
\frac{\mathbf{x}^H\mathbf{B}^H\mathbf{A}\mathbf{B}\mathbf{x}}{\mathbf{x}^H\mathbf{x}}=\mathop
 {\min }\limits_\mathbb{U} \mathop {\max }\limits_{\mathbf{x} \in \mathbb{F}}
\frac{\mathbf{x}^H\mathbf{A}\mathbf{x}}{\mathbf{x}^H\mathbf{x}}.
\end{equation}
Then we have $\mu_k = \mathop {\max }\limits_\mathbb{U} \mathop {\min }\limits_{\mathbf{x} \in \mathbb{F}} \frac{\mathbf{x}^H\mathbf{x}}{\mathbf{x}^H\mathbf{A}\mathbf{x}}$. Since $\mathbf{A}$ is positive definite, by Jensen's inequality, $\frac{\mathbf{x}^H\mathbf{x}}{\mathbf{x}^H\mathbf{A}\mathbf{x}}\le \frac{\mathbf{x}^H\mathbf{A}^{-1}\mathbf{x}}{\mathbf{x}^H\mathbf{x}}$ holds for any non-zero vector $\mathbf{x}$. Thus, the proof is completed. \hfill $\square$

Then, denoting $\left(\mathbf{I}_{{N}_{\mathrm{t}}} + \frac{1}{\sigma^2w_kN_\mathrm{t}} \mathbf{H}_{\mathrm{1},k}\mathbf{H}^H_{\mathrm{1},k}\right)$ as $\mathbf{A}_{k}$, and using Lemma \ref{Lemma:UB}, the objective in (\ref{Pro:opt-EVD-broadband}) can be further upper bounded as
\begin{equation}\label{evd:upperbound}
\begin{aligned}
J(\mathbf{V}_\mathrm{RF}) &\mathop  = \limits^{(a)}  \sum_{k = 0}^{N - 1}
\mathrm{tr}\left({\left(\mathbf{V}_\mathrm{RF}^H
\mathbf{A}_{k}\mathbf{V}_\mathrm{RF}\right)^{-1}}\right) \\&\mathop  \le
\limits^{(b)} \sum_{k = 0}^{N - 1}
\mathrm{tr}\left(\mathbf{V}_\mathrm{RF}^H
\mathbf{A}_{k}^{-1}\mathbf{V}_\mathrm{RF}\right) =
\mathrm{tr}(\mathbf{V}_\mathrm{RF}^H (\sum_{k = 0}^{N - 1}
\mathbf{A}_{k}^{-1})\mathbf{V}_\mathrm{RF})
\end{aligned}
\end{equation}
where (a) follows from the relationship between a matrix's trace and eigenvalues and (b) follows from Lemma \ref{Lemma:UB}.  Using the matrix inversion lemma \cite{1990Matrix}, we have
$\mathbf{A}_k^{-1} = \mathbf{I}_{N_\mathrm{t}} - \mathbf{G}_k$, where
$\mathbf{G}_k\triangleq\frac{1}{\sigma^2w_kN_\mathrm{t}}\mathbf{H}_{\mathrm{1},k}(\mathbf{I}_{N_\mathrm{s}} + \frac{1}{\sigma^2w_kN_\mathrm{t}}\mathbf{H}_{\mathrm{1},k}^H\mathbf{H}_{\mathrm{1},k})^{-1}\mathbf{H}_{\mathrm{1},k}^H$. The optimization problem in (\ref{Pro:opt-EVD-broadband}) can be converted to the one minimizing its upper bound in (\ref{evd:upperbound}), or equivalently
\begin{equation}\label{pro:opt-upper-broadband-approx}
\begin{array}{cl}
\displaystyle{\maximize_{\mathbf{V}_\mathrm{RF}}} & \mathrm{tr}\left(\mathbf{V}_\mathrm{RF}^H\left(\sum_{k = 0}^{N-1}\mathbf{G}_k\right)\mathbf{V}_\mathrm{RF}\right) \\
\mathrm{subject \; to} & \mathbf{V}_\mathrm{RF}^H\mathbf{V}_\mathrm{RF}={N_\mathrm{t}}\mathbf{I}_{N_\mathrm{RF}}.
\end{array}
\end{equation}
Similar to that for (\ref{pro:opt-low-broadband-approx}), the solution can be obtained through the EVD and phase extraction operations. To distinguish, we refer to this algorithm as the EVD-UB-HBF algorithm, where UB denotes the abbreviation for upper bound.}
%According to  \cite{KFAN},  the optimal $\mathbf{V}_\mathrm{RF}$ is
%$\sqrt{N_\mathrm{t}}$ times the isometric matrix containing the $N_\mathrm{RF}$
%eigenvectors associated with the largest $N_\mathrm{RF}$  eigenvalues of
%$\left(\sum_{k = 0}^{N-1}\mathbf{G}_k\right)$, then similar EVD and phase extraction operations in the EVD-LB-HBF can be applied to form the analog beamformers. in order to distinguish, we refer to this algorithm as EVD-UB-HBF.

\subsubsection{Analog Precoding Based on the OMP Method}\label{subsubsec:omp-low-com-broadband}
By combining the OMP-MMSE-HBF algorithm for narrowband multiuser mmWave MIMO systems \cite{OMP2016} and the OMP-based HBF algorithm aiming at maximizing the spectral efficiency for broadband mmWave MIMO systems \cite{OMP2014}, we come up with the low-complexity OMP-MMSE precoding algorithm for broadband mmWave systems. Specifically, by restricting the search range of $\mathbf{V}_\mathrm{RF}$ within a set of $N_{\mathrm{C}} \times N_{\mathrm{R}}$ basis vectors $\{{\mathbf{a}_\mathrm{t}} {(\theta_{1,1}^\mathrm{t})},\dots,{\mathbf{a}_\mathrm{t}} {(\theta_{N_\mathrm{C},N_\mathrm{R}}^\mathrm{t})}\}$, the hybrid beamforming problem can be rewritten as
\begin{equation}\label{pro:omp}
\begin{array}{cl}
\displaystyle{\minimize_{{\widehat {\mathbf{V}}}_{\mathrm{U},k},\beta_k}} & \sum_{k = 0}^{N - 1} \left({\left\| {\mathbf{I}_{N_\mathrm{s}} - \mathbf{H}_{1,k}\mathbf{A}_\mathrm{t}{{\widehat {\mathbf{V}}}_{\mathrm{U},k}}} \right\|} _F^2 + {\sigma ^2}\beta_k^2w_k\right)  \\
\mathrm{subject \; to} & {\left\| {\mathrm{diag}\{{\widehat {\mathbf{V}}}_{\mathrm{U},k}{\widehat {\mathbf{V}}}_{\mathrm{U},k}^H \} } \right\|_0} = {N_\mathrm{RF}}, \quad  \\&\mathrm{tr}({\mathbf{A}_\mathrm{t}}^H{\mathbf{A}_\mathrm{t}}{\widehat {\mathbf{V}}}_{\mathrm{U},k}{\widehat {\mathbf{V}}}_{\mathrm{U},k}^H)\le \beta_k^{-2},
\end{array}
\end{equation}
where $||.||_0$ denotes the zero norm of a matrix.  $\mathbf{A}_\mathrm{t}\triangleq[{\mathbf{a}_\mathrm{t}} {(\theta_{1,1}^\mathrm{t})},\dots,{\mathbf{a}_\mathrm{t}} {(\theta_{N_\mathrm{C},N_\mathrm{R}}^\mathrm{t})}]$ and ${\widehat {\mathbf{V}}}_{\mathrm{U},k}$ is an $N_{\mathrm{C}}N_{\mathrm{R}} \times N_\mathrm{s}$ matrix having $N_\mathrm{RF}$ non-zero rows which constitute $\mathbf{V}_{\mathrm{U},k}$ as defined in \cite{Heath2014,OMP2016,OMP2014}. With the readily derived closed-form solution of the digital precoder in (\ref{eqn:VD-k-broad}), the algorithm developed based on the OMP method can be applied to choose the columns of $\mathbf{A}_\mathrm{t}$ that are most strongly correlated with the residual error $\{\mathbf{V}_{\mathrm{RES},k}\}$ to form the analog precoder. % The detail procedure of the  OMP-MMSE-HBF algorithm can be found in \cite{OMP2016,OMP2014} and is limited for the space limitation .

{\section{Extend to Spectral Efficiency Based on the WMMSE Criterion}\label{sec:WMMSE}
Spectral efficiency is another important performance metric for the HBF design \cite{Heath2014,ZhangJun2016,YuWei2017}. {\color{black}Based on the full-digital beamforming design approach in \cite{WMMSE}, the authors in \cite{Heath2017} investigated the HBF design with the WMMSE criterion and connected it to the one for sum-rate maximization. However, their HBF algorithm is based on the OMP method, which has a limited feasible set for the analog beamformers, and is only for the narrowband scenario. In this section, following the design approach in \cite{WMMSE,Heath2017}, we first show that in the narrowband scenario our proposed HBF algorithms in Section \ref{sec:Design-Narrowband} can be extended to the ones for achieving better spectral efficiency than the OMP based algorithm. We also extend the sum-MSE minimization problem to the WMMSE problem and connect it to the spectral efficiency maximization problem in the broadband scenario. It is shown that the proposed broadband HBF algorithms in Section \ref{sec:design-broadband} can be generalized to the ones for maximizing the spectral efficiency. Simulation results in Section \ref{sec:simulation} will show that the WMMSE HBF algorithms proposed in this section provide better or comparable spectral efficiency than the conventional ones \cite{ZhangJun2016,YuWei2016,YuWei2017,Heath2017}.}

First start from the narrowband scenario. Assuming that the transmitted symbols follow a Gaussian distribution, the achievable spectral efficiency is then given by $R = \mathrm{log}\; \mathrm{det}(\mathbf{I}_{N_\mathrm{s}} + \frac{1}{\sigma^2}( ( \mathbf{W}^H\mathbf{W}) ^{-1}\mathbf{W}^H) \mathbf{H}\mathbf{V}\mathbf{V}^H\mathbf{H}^H\mathbf{W})$,	where $\mathbf{V}=\mathbf{V}_\mathrm{RF}\mathbf{V}_\mathrm{B}$ and $\mathbf{W}=\mathbf{W}_\mathrm{RF}\mathbf{W}_\mathrm{B}$ denote the hybrid precoder and combiner, respectively. Inspired by \cite{WMMSE,Heath2017}, a suboptimal but efficient HBF design for maximizing the spectral efficiency can be connected to the following WMMSE problem
%\footnote{{It is worth noting that although in the full-digital beamforming design the equivalence between the spectral efficiency maximization problem and the WMMSE problem can be strictly proved, it cannot be proved in the HBF design due to the constant modulus constraint on the analog beamformers. Nevertheless, this problem conversion  provides an efficient approach to solve the spectral efficiency maximization problem, as will be demonstrated via simulations in Section \ref{sec:simulation}.}}
 \begin{equation}\label{pro:opt-WMMSE}
\begin{array}{cl}
\displaystyle{\minimize_{{\mathbf{V}}, {\mathbf{W}}, \mathbf{\Lambda}, \beta}} & \mathrm{tr}(\mathbf{\Lambda}\mathbf{T}) - \mathrm{log}|\mathbf{\Lambda}| \\
\mathrm{subject \; to} & \|{\mathbf{V}}\|_F^2\leq 1;\quad |[{\mathbf{V}}_{\mathrm{RF}}] _{ij}|^2=1, \forall i,j;  \quad\\&|[{\mathbf{W}}_{\mathrm{RF}}] _{ml}|^2=1, \forall m,l,
\end{array}
\end{equation}
where $\mathbf{\Lambda}$ is an $N_\mathrm{s}\times N_\mathrm{s}$ weighting  matrix  to be optimized, and $\mathbf{T}\triangleq E\{\left(\beta^{-1}\mathbf{y}-\mathbf{s}\right)\left(\beta^{-1}\mathbf{y}-\mathbf{s}\right)^H\}$ denotes the modified MSE matrix. According to \cite{WMMSE,Heath2017}, a three-step procedure is applied to solve (\ref{pro:opt-WMMSE}). In the first step, $\mathbf{W}$ is optimized by fixing $\mathbf{\Lambda}$ and $\mathbf{V}$ in (\ref{pro:opt-WMMSE}). That is,
\begin{equation}\label{pro:WMSEopt-combining}
\begin{array}{cl}
\displaystyle{\minimize_{\mathbf{W}_\mathrm{RF},{\mathbf{W}}_\mathrm{B}}} & \mathrm{tr}( 	\mathbf{\Lambda}({\mathbf{W}}^H\mathbf{H}_\mathrm{2}{\mathbf{H}}_\mathrm{2}^H{\mathbf{W}} -{\mathbf{W}}^H \mathbf{H}_\mathrm{2} - \mathbf{H}_\mathrm{2}^H{\mathbf{W}} \\&\quad+{\sigma^2}\beta^{-2}{\mathbf{W}}^H{\mathbf{W}}+\mathbf{I}_{N_\mathrm{s}} ))\\
\mathrm{subject \; to} & \quad|[\mathbf{W}_{\mathrm{RF}}]_{ml}|=1, \forall m,l,
\end{array}
\end{equation}
where $\mathbf{H}_\mathrm{2} \triangleq \mathbf{H}\mathbf{V}_\mathrm{RF}{{\mathbf{V}}_\mathrm{U}}$. By comparing (\ref{pro:WMSEopt-combining}) with (\ref{pro:opt-combining}), it can be shown that the optimal $\mathbf{W}_\mathrm{B}$ has exactly the same form as that in (\ref{eqn:WD}).  After substituting the optimal $\mathbf{W}_\mathrm{B}$ into (\ref{pro:WMSEopt-combining}), the objective function for $\mathbf{W}_\mathrm{RF}$ is given by
\begin{equation}\label{newI}
\begin{aligned}
I(\mathbf{W}_\mathrm{RF}) &\triangleq\mathrm{tr}(\mathbf{\Lambda}(\mathbf{I}_{{N}_{\mathrm{s}}} + \sigma^{-2}\beta^{2} \mathbf{H}_\mathrm{2}^H\mathbf{W}_\mathrm{RF}\\&\quad\quad\times{\left(\mathbf{W}_\mathrm{RF}^H {\mathbf{W}_\mathrm{RF}}\right)^{ - 1}}\mathbf{W}_\mathrm{RF}^H\mathbf{H}_\mathrm{2}^H)^{ - 1}),
\end{aligned}
\end{equation}
which has almost the same form as (\ref{pro:opt-W-RF}) except a constant matrix multiplier $\mathbf{\Lambda}$. Thus, both the MO-HBF and the GEVD-HBF algorithms in Section \ref{sec:Design-Narrowband} can be modified to solve the new problem. In the second step, the weighting  matrix $\mathbf{\Lambda}$ is optimized with fixed $\mathbf{W}$ and $\mathbf{V}$. By differentiating the objective function in (\ref{pro:opt-WMMSE}) with respect to $\mathbf{\Lambda}$ and then setting the result to zero, the optimal  $\mathbf{\Lambda}$ is then given by $\mathbf{\Lambda} = \mathbf{T}^{-1}$. In the last step, $\mathbf{V}$ is optimized through the following problem with the newly updated $\mathbf{W}$ and $\mathbf{\Lambda}$.
\begin{equation}\label{pro:Vopt-precoding}
\begin{array}{cl}
\displaystyle{\minimize_{\mathbf{V}_\mathrm{RF},\mathbf{V}_\mathrm{U},\beta}} & \mathrm{tr} 	(\mathbf{\Lambda}(\mathbf{H}_\mathrm{1}^H\mathbf{V}\mathbf{V}^H\mathbf{H}_\mathrm{1} - \mathbf{H}_\mathrm{1}^H\mathbf{V} -\mathbf{V}^H \mathbf{H}_\mathrm{1} \\&\quad + {\sigma^2}\beta^{-2} {\mathbf{W}}^H{\mathbf{W}} +\mathbf{I}_{N_\mathrm{s}}) )\\
\mathrm{subject \; to} & \|{\mathbf{V}}\|_F^2\leq 1;\quad|[\mathbf{V}_{\mathrm{RF}}]_{ij}|=1, \forall i,j,
\end{array}
\end{equation}
where $\mathbf{H}_\mathrm{1}\triangleq\mathbf{H}^H \mathbf{W}_\mathrm{RF}\mathbf{W}_\mathrm{B}$. Then, the optimal $\mathbf{V}_\mathrm{U}$ and $\beta$ are given by
\begin{eqnarray}\label{WMMSE_conv}\nonumber
&\beta =\left(\mathrm{tr}\left(\mathbf{V}_\mathrm{RF} {{\mathbf{V}}_\mathrm{U}} {{\mathbf{V}}^H_\mathrm{U}}\mathbf{V}_\mathrm{RF}^H\right)\right)^{-\frac{1}{2}}, \\
&{{\mathbf{V}}_\mathrm{U}} =(\mathbf{V}_\mathrm{RF}^H\mathbf{H}_\mathrm{1}\mathbf{\Lambda}\mathbf{H}_\mathrm{1}^H\mathbf{V}_\mathrm{RF}+{\sigma^2}\psi \mathbf{V}_\mathrm{RF}^H\mathbf{V}_\mathrm{RF})^{-1}\mathbf{V}_\mathrm{RF}^H \mathbf{H}_\mathrm{1}\mathbf{\Lambda},\nonumber
\end{eqnarray}
where $\psi\triangleq \mathrm{tr}(\mathbf{\Lambda}\mathbf{W}^H\mathbf{W})$ is a constant scalar during the optimization for $\mathbf{V}$. Substituting the optimal $\mathbf{V}_\mathrm{U}$ and $\beta$ into the objective function in (\ref{pro:Vopt-precoding}), we have
\begin{equation}\label{newJ}
J(\mathbf{V}_\mathrm{RF}) \triangleq\mathrm{tr}({(\mathbf{\Lambda}^{-1} + \frac{1}{\sigma^2\psi} \mathbf{H}_\mathrm{1}^H\mathbf{V}_\mathrm{RF}{(\mathbf{V}_\mathrm{RF}^H {\mathbf{V}_\mathrm{RF}})^{ - 1}}\mathbf{V}_\mathrm{RF}^H\mathbf{H}_\mathrm{1}^H)^{ - 1}}).
\end{equation}
By comparing it with (\ref{eqn:MSE-as-fun-VRF}), we  see that they have the same form except that the constant identity matrix $\mathbf{I}_{N_\mathrm{s}}$ in (\ref{eqn:MSE-as-fun-VRF}) is replaced by another constant matrix $\mathbf{\Lambda}$ in (\ref{newJ}). Thus, both the MO-HBF and the GEVD-HBF algorithms in Section \ref{sec:Design-Narrowband} can also be applied here. By iteratively performing the above three steps, the optimization problem in (\ref{pro:opt-WMMSE}) can finally be solved.

Following the above design approach, we now consider the the broadband scenario. From (\ref{pro:opt-WMMSE}), we formulate the following broadband WMMSE HBF optimization problem
\begin{equation}\label{pro:opt-WMMSE-broadband}
\begin{array}{cl}
\displaystyle{\minimize_{{\mathbf{V}_k}, {\mathbf{W}_k}, \mathbf{\Lambda}_k,\beta_k}} & \sum\limits_{k=0}^{N-1}\left(\mathrm{tr}(\mathbf{\Lambda}_k\mathbf{T}_k) - \mathrm{log}|\mathbf{\Lambda}_k|\right) \\
\mathrm{subject \; to} & \|{\mathbf{V}_k}\|_F^2\leq 1; \\& |[{\mathbf{V}}_{\mathrm{RF}}] _{ij}|^2=1, \forall i,j;  \quad \\&|[{\mathbf{W}}_{\mathrm{RF}}] _{ml}|^2=1, \forall m,l,
\end{array}
\end{equation}
where $\mathbf{\Lambda}_k$ and $\mathbf{T}_k$ denote the weighting matrix and
the MSE-matrix at the $k$th subcarrier, respectively. By combining the procedure in Section \ref{sec:design-broadband} and that for solving (\ref{pro:opt-WMMSE}), we can solve the WMMSE problem in (\ref{pro:opt-WMMSE-broadband}). In particular, it can be shown that the MO-HBF and the EVD-LB-HBF algorithms can be directly applied to solve the problem with slight modification as the only difference is the constant matrix $\mathbf{\Lambda}_k$. However, the EVD-UB-HBF algorithm cannot be generalized for the WMMSE problem as Lemma \ref{Lemma:UB} does not hold with the weighting matrices. Note that compared with the conventional algorithms \cite{ZhangJun2016, YuWei2016,YuWei2017}, our proposed WMMSE based HBF algorithms can benefit from the alternating optimization between the transmitter and receiver sub-problems, and thus possess competitive performance, as will be shown in Section \ref{sec:simulation}.} %is no longer available to derive the upper bound in this formulation.} %Nevertheless, simulation results in Section VII will  show that the EVD-HBF algorithm using the MMSE criterion can also achieve good spectral efficiency performance, and be regarded as an alternative low-complexity approach for the MO-HBF algorithm.

\section{System Evaluation}\label{sec:evaluation}
{\color{black}In this section, we first discuss the convergence property for all the proposed HBF algorithms and then analyze their computational complexity.

\subsection{Convergence}\label{subsec:convergence}
It is worth noting that all the proposed HBF algorithms share the same design procedure in the optimization of the digital beamformers, where the optimal digital precoder or combiner has a closed-form solution obtained via the KKT conditions. Thus, for given analog beamformers, the optimization step of the digital beamformers always ensures the decrease of the objective function \cite{Boyd2004}. Therefore, the convergence of each HBF algorithm depends on its optimization step for the analog beamformer, which is discussed as follows.
\begin{itemize}

\item \underline{MO-HBF:} In this algorithm, the analog beamformers are optimized via the MO method. According to Theorem 4.3.1 in \cite{MO2009}, the algorithm using the MO method is guaranteed to converge to the point where the gradient of the objective function is zero \cite{ZhangJun2016}. Therefore, each step of the whole alternating MO-HBF algorithm ensures the decrease of the objective function and the convergence can be strictly proved.

\item \underline{GEVD-HBF:} Unlike the MO-HBF algorithm, the convergence of the GEVD-HBF algorithm in Section \ref{sec:Design-Narrowband} cannot be strictly proved. This is because its derivations are based on the approximations of $\mathbf{V}_\mathrm{RF}^H\mathbf{V}_\mathrm{RF} \approx \mathbf{I}_{N_\mathrm{RF}}$ and $\mathbf{W}_\mathrm{RF}^H\mathbf{W}_\mathrm{RF} \approx \mathbf{I}_{N_\mathrm{RF}}$, and the phase extraction operation further raises the difficulty. Nevertheless, simulation results in Section \ref{sec:simulation} will show that the whole alternating GEVD-HBF algorithm has fast convergence. An intuitive explanation is that excluding the orthogonal approximation and the phase extraction operation, other steps in the algorithm ensure the strict decrease of the objective function, and the orthogonal approximations on the analog beamforming matrices and the phase extraction operation have no great impact on the changing trend of the objective function.
	
\item \underline{EVD-UB-HBF, EVD-LB-HBF and OMP-MMSE-HBF:} For the EVD-UB-HBF and EVD-LB-HBF algorithms in Section \ref{subsubsec:low-com-broadband}, the monotonic decrease of the original objective function is not ensured when optimizing the analog beamformers due to the orthogonal approximation, the lower or upper bounding operation, and the phase extraction operation. For the OMP-MMSE-HBF algorithm in Section \ref{subsubsec:omp-low-com-broadband}, the performance loss due to the limitation of the feasible set of the analog beamformers cannot guarantee the strict convergence. Nevertheless, simulation results in Section \ref{sec:simulation} will show that these algorithms converge in most cases.
	
\item \underline{WMMSE algorithms:} The above discussion on the convergence of the five HBF algorithms using the MMSE criterion can be extended to their counterparts with the WMMSE criterion in Section \ref{sec:WMMSE}. The main difference is the additional step for optimizing the weighting matrix in the WMMSE based algorithms. As shown in Section \ref{sec:WMMSE}, the optimal weighting matrix has a closed-form solution via the KKT conditions, which ensures the decrease of the objective function \cite{Boyd2004}. Thus, the convergence depends on the design of the analog beamformers. As shown in (\ref{newI}) and (\ref{newJ}), since the weighting matrix is regarded as a constant matrix in the optimization steps for the analog beamformers, following the above discussion on the convergence of the MMSE based HBF algorithms, it can be concluded that with the WMMSE criterion, the convergence of the MO-HBF algorithm can be strictly proved, and that of other algorithms cannot be proved in spite of the observation of convergence from simulation results.
	
\end{itemize}
}

\subsection{Complexity analysis}\label{sec:complexity}
{In this subsection we analyze the computational complexity in terms of the number of complex multiplications for all the proposed MMSE HBF algorithms. The complexity of the WMMSE based algorithms can be regarded as the same as that of the MMSE based counterparts, as the dimension of the weighting matrix $\mathbf{\Lambda}$ is only $N_\mathrm{s}\times N_\mathrm{s}$ and the related additional complexity is negligible. Since it has been shown that the transmit precoding and receive combining sub-problems can be solved in the same procedure, we take the transmit precoding as an example for complexity analysis. Furthermore, we focus on the complexity in computing the analog precoder and ignore that in the digital one. This is because all the proposed HBF algorithms have the same complexity in computing the digital one, which is also much less than that in the analog one due to the difference between their matrices' sizes. For simplification, we denote $N_\mathrm{ant} = \mathrm{max}\{N_\mathrm{t},N_\mathrm{r}\}$ and assume $N_\mathrm{RF} = N_\mathrm{s}$.

\subsubsection{Narrowband algorithms} For the MO-HBF algorithm, the main complexity in each inner iteration includes  the following three parts:
\begin{itemize}
\item \underline{Computation of the conjugate gradient}: According to (\ref{eqn:gradient-J}), the total complexity in computing the gradient is $(4N^2_\mathrm{ant}N_\mathrm{RF} + 7N_\mathrm{ant}N^2_\mathrm{RF}+2N^3_\mathrm{RF} + 2\mathcal{O}(N_\mathrm{RF}^3))$, where $2\mathcal{O}(N_\mathrm{RF}^3)$ results from the inversion of two $N_\mathrm{RF} \times N_\mathrm{RF}$ matrices.

\item \underline{Orthogonal projection and retraction operations}: According to \cite{ZhangJun2016}, the orthogonal projection is essentially the Hadamard production which takes $2N_\mathrm{ant}N_\mathrm{RF}$ multiplications. In addition, the complexity of the retraction operation is $N_\mathrm{ant}N_\mathrm{RF}$.

\item \underline{Line search}: To guarantee the convergence, we utilize the well-known Armijo backtracking line search, whose complexity is $(6N_\mathrm{ant}N^2_\mathrm{RF} + 2N_\mathrm{RF}^3 + 2\mathcal{O}(N_\mathrm{RF}^3))$, where $2\mathcal{O}(N_\mathrm{RF}^3)$ results from the inversion of  two $N_\mathrm{RF} \times N_\mathrm{RF}$ matrices.
\end{itemize}
Denote the numbers of the inner and outer iterations as $N_\mathrm{in}$ and $N_\mathrm{out}$ respectively, the total complexity of MO-HBF is $N_\mathrm{out}N_\mathrm{in}(4N^2_\mathrm{ant}N_\mathrm{RF}+ 13N_\mathrm{ant}N^2_\mathrm{RF} + 3N_\mathrm{ant}N_\mathrm{RF} + 4N_\mathrm{RF}^3 + 4\mathcal{O}(N_\mathrm{RF}^3))$.

For the GEVD-HBF algorithm,  the main complexity includes the following two parts:
\begin{itemize}
\item \underline{Before the GEVD operation}: The complexity in computing $\mathbf{A}_{m}$, $\mathbf{U}_{m}$ and $\mathbf{W}_{m}$ is $(2N_\mathrm{ant}^2N_\mathrm{RF} + 5N_\mathrm{RF}^2N_\mathrm{ant} + 2N_\mathrm{RF}^3 + \mathcal{O}(N_\mathrm{RF}^3))$, where $\mathcal{O}(N_\mathrm{RF}^3)$ represents the complexity of the inversion of an $N_\mathrm{RF} \times N_\mathrm{RF}$ matrix.
\item \underline{The GEVD operation}: The complexity of the GEVD operation is in the order of $\mathcal{O}(N_\mathrm{ant}^3)$. However, as only the largest generalized eigenvector needs to be computed, the complexity can be reduced to $\mathcal{O}(N_\mathrm{p}N_\mathrm{ant}^2)$ by using the power method \cite{powermethod}, where $N_\mathrm{p}$ denotes the number of iterations in the power method. By extensive simulations, it is observed that $N_\mathrm{p}=10$ would be large enough to obtain an accurate result.
  \end{itemize}
According to the description of the iteration stop condition in Section \ref{stopping condition}, one inner iteration (i.e., $N_\mathrm{in}=1$) is enough for GEVD-HBF. Then, the total complexity of GEVD-HBF is $N_\mathrm{out}(\mathcal{O}(N_\mathrm{p}N_\mathrm{ant}^2) + \mathcal{O}(N_\mathrm{RF}^3) + 2N_\mathrm{ant}^2N_\mathrm{RF} + 5N_\mathrm{RF}^2N_\mathrm{ant} + 2N_\mathrm{RF}^3)$.

\subsubsection{Broadband algorithms}
For the MO-HBF algorithm in the broadband scenario, according  to (\ref{eqn:gW-broadband}), the complexity is approximated as $N$ times that in the narrowband scenario, which is $NN_\mathrm{out}N_\mathrm{in}(4N^2_\mathrm{ant}N_\mathrm{RF}+ 13N_\mathrm{ant}N^2_\mathrm{RF} + 3N_\mathrm{ant}N_\mathrm{RF} + 4N_\mathrm{RF}^3 + 4\mathcal{O}(N_\mathrm{RF}^3))$. For the EVD-LB-HBF algorithm, the main complexity is in computing  $(\sum_{k = 0}^{N - 1}\mathbf{H}_{\mathrm{1},k}\mathbf{H}_{\mathrm{1},k}^H)$ and the EVD operation. The former one is $(NN_\mathrm{ant}^2N_\mathrm{RF})$. For the latter one, according to the power method \cite{powermethod2}, the complexity can be reduced to $\mathcal{O}(N_\mathrm{RF}N_\mathrm{ant}^2)$ for computing the largest $N_\mathrm{RF}$ eigenvectors. As for the EVD-UB-HBF algorithm, while the analysis of the EVD operation is similar, the complexity in computing $(\sum_{k = 0}^{N-1}\mathbf{G}_k)$ is $(NN_\mathrm{ant}^2N_\mathrm{RF} + 2NN_\mathrm{ant}N_\mathrm{RF}^2+\mathcal{O}(N_\mathrm{RF}^3))$. Finally, the main complexity of OMP-MMSE-HBF algorithm is in computing $(\sum_{k = 0}^{N - 1}{\mathbf{\Phi}_k \mathbf{\Phi}_k ^H})$, which is $NN_\mathrm{out}(N_\mathrm{RF}N_\mathrm{s}N_\mathrm{C}N_\mathrm{R}N_\mathrm{ant}+ N_\mathrm{RF}N_\mathrm{C}^2N_\mathrm{R}^2)$.

In summary, we list the above complexity evaluation results in Table \ref{tab:complexity_specific}.  For a more intuitive expression, we provide the average numbers of inner and outer iterations over 1000 independent channel realizations in simulations, where the mmWave MIMO system configuration is given in Section \ref{sec:simulation} and the iteration stop conditions are set according to Section \ref{stopping condition}. It can be seen from Table \ref{tab:complexity_specific} that all the HBF algorithms have similar number of outer iterations. However, because the MO-HBF algorithm requires a large number of inner iterations in the gradient descent operation, it has the highest computational complexity.}

{
\begin{table*}[]
	\centering
	{ \caption{Computational Complexity for Different HBF Algorithms} \label{tab:complexity_specific}}
{
	\begin{tabular}{|c|c|c|c|}
		\hline
		\multicolumn{4}{|c|}{Narrowband Scenario} \\ \hline\hline
		HBF Algorithms & Computational Complexity & $N_\mathrm{out }$ &$N_\mathrm{in}$\\ \hline
		MO-HBF &  $N_\mathrm{out}N_\mathrm{in}(4N^2_\mathrm{ant}N_\mathrm{RF}+ 13N_\mathrm{ant}N^2_\mathrm{RF} + 3N_\mathrm{ant}N_\mathrm{RF} + 4N_\mathrm{RF}^3 + 4\mathcal{O}(N_\mathrm{RF}^3))$ & 4.8 & 49.8\\ \hline
		GEVD-HBF & $N_\mathrm{out}(\mathcal{O}(N_\mathrm{ant}^2N_\mathrm{p}) + 2N_\mathrm{ant}^2N_\mathrm{RF} + 5N_\mathrm{RF}^2N_\mathrm{ant} +   2N_\mathrm{RF}^3 + \mathcal{O}(N_\mathrm{RF}^3))$ & 5.8 &\diagbox{}{}
		
	 \\ \hline\hline
		\multicolumn{4}{|c|}{Broadband Scenario} \\ \hline\hline
		HBF Algorithms & Computational Complexity & $N_\mathrm{out}$ &$N_\mathrm{in}$ \\ \hline
		MO-HBF & $NN_\mathrm{out}N_\mathrm{in}(4N^2_\mathrm{ant}N_\mathrm{RF}+ 13N_\mathrm{ant}N^2_\mathrm{RF} + 3N_\mathrm{ant}N_\mathrm{RF} + 4N_\mathrm{RF}^3 + 4\mathcal{O}(N_\mathrm{RF}^3))$ & 4.7 & 52.3\\ \hline
		EVD-LB-HBF & $N_\mathrm{out}(\mathcal{O}(N_\mathrm{ant}^2N_\mathrm{RF}) +  NN_\mathrm{ant}^2N_\mathrm{RF})$ & 5.9 &\diagbox{}{}\\ \hline
			EVD-UB-HBF & $N_\mathrm{out}(\mathcal{O}(N_\mathrm{ant}^2N_\mathrm{RF}) + NN_\mathrm{ant}^2N_\mathrm{RF}  +2NN_\mathrm{ant}N_\mathrm{RF}^2  + \mathcal{O}(N_\mathrm{RF}^3))$ & 5.7 &\diagbox{}{}\\ \hline
		OMP-MMSE-HBF & $NN_\mathrm{out}(N_\mathrm{RF}N_\mathrm{s}N_\mathrm{C}N_\mathrm{R}N_\mathrm{ant}+ N_\mathrm{RF}N_\mathrm{C}^2N_\mathrm{R}^2)$ & 6.2 &\diagbox{}{}\\ \hline
\end{tabular}
}
\end{table*}
}
%\begin{figure}
%	\begin{minipage}[c]{0.45\linewidth}
%		\centering
%		\includegraphics[width=2.8in]{convergenceNB.eps}
%		\caption{Average MSE performance v.s. the number of outer iterations when $\mathrm{SNR} = -16$dB in the narrowband scenario.}
%		\label{fig:side:a}
%	\end{minipage}%
%	\hspace{2.2cm}
%	\begin{minipage}[c]{0.45\linewidth}
%		\centering
%		\includegraphics[width=2.8in]{convergenceNB.eps}
%		\caption{Average MSE performance v.s. the number of outer iterations when $\mathrm{SNR} = -16$dB in the narrowband scenario.}
%		\label{fig:side:b}
%	\end{minipage}
%\end{figure}

\section{Simulation Results}\label{sec:simulation}
In this section, simulation results are provided to show the performance of the proposed HBF algorithms, in comparison with existing HBF algorithms and the optimal full-digital algorithms based on the MMSE criterion. The channel models for both the narrowband and the broadband scenarios, as introduced in Section \ref{subsec:system-model} and Section \ref{subsec:system-model-broadband}, respectively, are used in the simulation, where the number of clusters is set to $N_\mathrm{C}=5$ and the number of rays in each clusters is set to $N_\mathrm{R}=10$. Similar to \cite{ZhangJun2016,YuWei2016}, we assume that $\alpha_{ij} \sim \mathcal{CN}(0,1)$ and the angles of arrival and departure are generated according to the Laplacian distribution with the mean cluster angles $\overline{\theta_i^\mathrm{r}}$ and $\overline {\theta_i^\mathrm{t}}$, which are independently and uniformly distributed in $[0,2\pi]$. The angular spread is 10 degrees within each cluster. It is assumed that the channel estimation and system synchronization are perfect. Throughout the simulation, the numbers of transmit and receive antennas are set to $N_\mathrm{t} = N_\mathrm{r} =64$ unless otherwise mentioned and uncoded quadrature phase shift keying (QPSK) modulation is considered. Besides, SNR is defined as $\frac{1}{\sigma^2}$. Unless otherwise stated, we assume that the HBF optimization starts from the transmit precoding optimization and apply the proposed VFD initialization method in Section \ref{Initialization} to generate a hybrid combiner in the simulation of the proposed HBF algorithms.

\subsection{Performance in the Narrowband Scenario}\label{subsec:sim-narrowband}
% \begin{figure*}
% 	\centering
% 	\includegraphics[width=2.5in]{convergenceNB.eps}
% 	\caption{Average MSE performance v.s. the number of outer iterations when $\mathrm{SNR} = -16$dB in the narrowband scenario.}
% 	\label{fig:MSEvsIt-Narrow}
% \end{figure*}
 \begin{figure}[!t]
 		\centering
 		\includegraphics[width=3.8in]{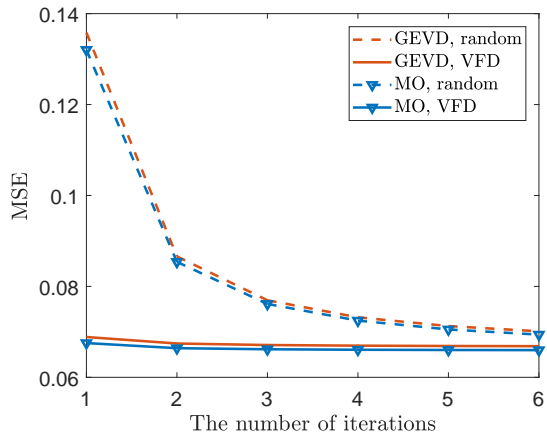}
 		\caption{Average MSE performance v.s. the number of outer iterations when $\mathrm{SNR} = -16$dB in the narrowband scenario.}
 		\label{fig:MSEvsIt-Narrow}
\end{figure}

 	 \begin{figure}[!t]
 		\centering
 		\includegraphics[width=3.8in]{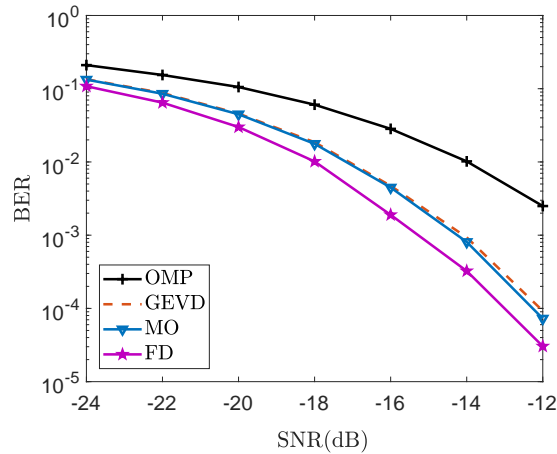}
 	\caption{BER v.s. SNR for different HBF algorithms when $N_\mathrm{RF} = N_\mathrm{s}=2$ in the narrowband scenario.}
 	    \label{fig:BERvsSNR-Narrow}
 	\end{figure}
To test the convergence and performance of the proposed algorithms, we first consider the narrowband scenario. Fig. \ref{fig:MSEvsIt-Narrow} shows the optimized MSE value averaged over 1000 channel realizations as a function of the number of  outer iterations  when $\mathrm{SNR}=-10\mathrm{dB}$ and $N_\mathrm{s}=N_\mathrm{RF}=2$. As shown in Section \ref{sec:evaluation}, the proposed MO-HBF algorithm (labeled with `MO') is guaranteed to converge. Although the proposed GEVD-HBF algorithm (labeled with `GEVD') does not necessarily experience a monotonic convergence, the simulation results show that this is often the case. Furthermore, the performance of two initialization methods, i.e., the random initiation method (labeled with `random') and the proposed VFD method in Section \ref{Initialization} (labeled with `VFD') are compared\footnote{{ Note that according to Section \ref{Initialization}, as in the proposed HBF algorithms with the VFD initialization at least one outer iteration is needed to obtain the hybrid beamformers at both two sides, the x-axis in the figure starts from `1'.}}. As shown in the figure, by using a virtual full-digital combiner (the optimal full-digital combiner in \cite{Smapth:2001}) as the initialization of the hybrid receive combiner in the proposed VFD  method, both algorithms quickly converge within a few outer iterations and even with some MSE improvement.
% \begin{figure*}[!t]
% 	\centering
% 	\begin{center}
% 		\centerline{\subfigure[]{\includegraphics[width=2.5in]{BERNB.eps}\label{fig:BERvsSNR-Narrow}} \hfil \subfigure[]{\includegraphics[width=2.5in]{rateNB.eps}\label{fig:narrow_rate}}\hfil }
% 		\caption{Performance comparison for different HBF algorithms in the narrowband scenario. (a) BER v.s. SNR. (b) Spectral efficiency v.s. SNR.} \label{fig:sim_NB}
% 	\end{center}
% \end{figure*}

   \begin{figure}[t]
 		\centering
 		\includegraphics[width=3.8in]{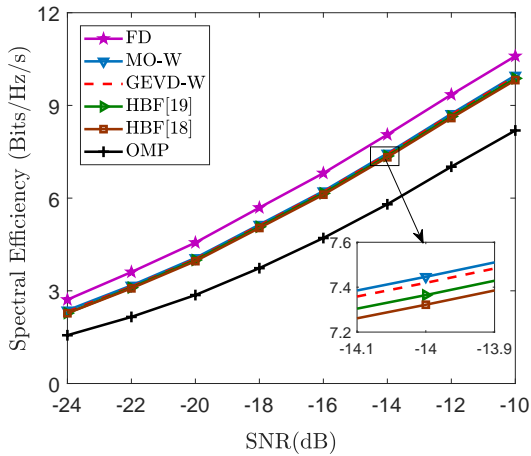}
 		\caption{Spectral efficiency v.s. SNR for different HBF algorithms when $N_\mathrm{RF} = N_\mathrm{s}=2$ in the narrowband scenario.}
 	\label{fig:narrow_rate}
 	\end{figure}%
 	\begin{figure}[t]
 		\centering
 		\includegraphics[width=3.8in]{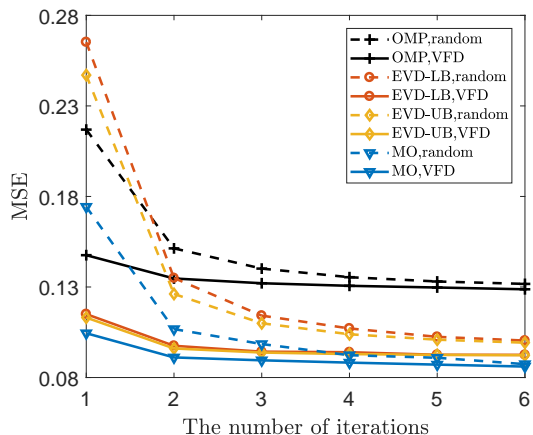}
 		\caption{Average MSE performance v.s. the number of  outer iterations when $\mathrm{SNR} = -10$dB, $N_\mathrm{RF}=N_\mathrm{s}=2$ in the broadband scenario.}
 		\label{convergenceWB}
 \end{figure}

Next, the BER performance of the proposed MO-HBF and GEVD-HBF algorithms (labeled with `MO' and `GEVD') in Section \ref{sec:Design-Narrowband} is presented. For comparison, the performance of the conventional OMP-based  algorithm in \cite{Lee2015,Heath2017} (labeled with `OMP') and that of the full-digital beamforming algorithm (labeled with `FD') based on the MMSE criterion are also demonstrated. Fig. \ref{fig:BERvsSNR-Narrow} shows the BER performance as a function of SNR for different algorithms with $N_\mathrm{s}=N_\mathrm{RF}=2$. It can be seen that the two proposed HBF algorithms  significantly outperform the conventional OMP-based algorithm and approach the full-digital one within 1dB. { This is because the OMP algorithm is limited to a predefined set consisting of the antenna array response vectors. Thus, the size of the feasible set is reduced greatly, which leads to the worst performance among all the algorithms.} The proposed GEVD-HBF algorithm performs closely to the MO-HBF algorithm and can be regarded as an alternative low-complexity algorithm.

{Fig. \ref{fig:narrow_rate} shows the spectral efficiency as a function of SNR for the two proposed narrowband HBF algorithms with the WMMSE criterion (labeled with 'MO-W' and 'GEVD-W') in Section \ref{sec:WMMSE}. For comparison, the performance of the full-digital beamforming algorithm in \cite{Heath2014} and the two conventional HBF algorithms in \cite{ZhangJun2016} and \cite{YuWei2016} (labeled with 'HBF \cite{ZhangJun2016}' and 'HBF \cite{YuWei2016}') aiming at maximizing the spectral efficiency is also provided. It can been seen that except the OMP algorithm, all the other HBF algorithms perform quite close to each other. The proposed WMMSE based algorithms perform slightly better than the conventional HBF algorithms. This is because as shown in Section \ref{sec:WMMSE} the optimization based on the WMMSE criterion with appropriate weights is an alternative approach to the maximum spectral efficiency objective, and the proposed WMMSE based HBF algorithms benefit from the alternating optimization between the transmit and receive sub-problems while the conventional ones cannot.}

\subsection{Performance in the Broadband Scenario}\label{subsec:sim-broadband}
% \begin{figure*}
% 	\centering
% 	\includegraphics[width=2.5in]{CVWB.eps}
% 	\caption{Average MSE performance v.s. the number of  outer iterations when $\mathrm{SNR} = -10$dB, $N_\mathrm{RF} = N_\mathrm{s} =2$ in the broadband scenario.}
% 	\label{convergenceWB}
% \end{figure*}

% \begin{figure*}[!t]
% 	\centering
% 	\begin{center}
% 		\centerline{\subfigure[]{\includegraphics[width=2.5in]{BERWB.eps}\label{ori_ber}} \hfil \subfigure[]{\includegraphics[width=2.5in]{ori_rate.eps}\label{ori_rate}}\hfil }
% 		\caption{Performance for different HBF algorithms when $N_\mathrm{RF} = N_\mathrm{s}=2$ in the broadband scenario. (a) BER v.s. SNR. (b) Spectral efficiency v.s. SNR.} \label{fig:sim_WBori}
% 	\end{center}
% \end{figure*}
%
 %\begin{figure*}[!t]
 %\centering
 %\includegraphics[width=3.7in]{BERWB.eps}%or 8.5cm
 %\caption{BER v.s. SNR for different HBF algorithms when  $N_\mathrm{RF} = N_\mathrm{s}=2$ in the broadband scenario.}
 %\label{ori_ber}
 %\end{figure*}
 %
 %\begin{figure*}
 %\centering
 %\includegraphics[width=3.7in]{ori_rate.eps}
 %\caption{ The spectral efficiency v.s. SNR for different HBF algorithmswhen  $N_\mathrm{RF} = N_\mathrm{s}=2$ in the broadband scenario.}
 %\label{ori_rate}
 %\end{figure*}
   \begin{figure}
 		\centering
 		\includegraphics[width=3.8in]{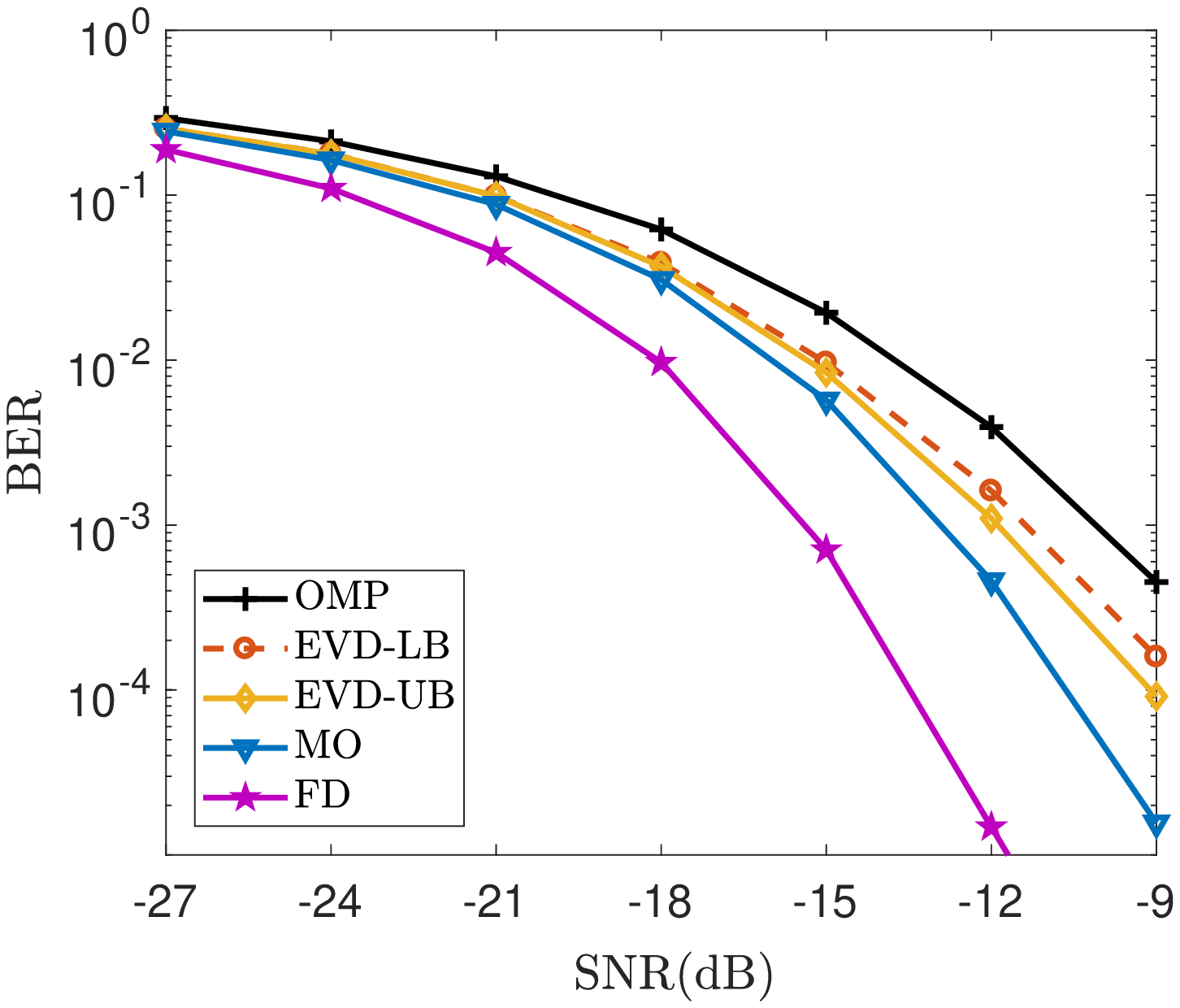}
 		\caption{BER v.s. SNR for different HBF algorithms when $N_\mathrm{RF} = N_\mathrm{s}=2$ in the broadband scenario.}
 		\label{ori_ber}
 	\end{figure}%
 	\begin{figure}
 		\centering
 		\includegraphics[width=3.8in]{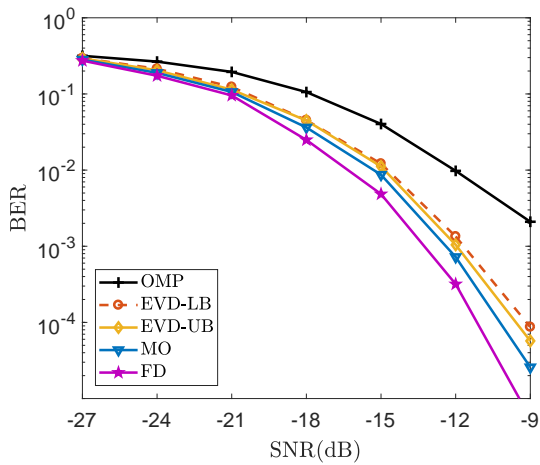}
 		\caption{BER v.s. SNR for different HBF algorithms when $N_\mathrm{t}=64, N_\mathrm{r}=32$, $N_\mathrm{RF}=4, N_\mathrm{s}=2$ in the broadband scenario.}
 		\label{yuwei_ber}
 	\end{figure}

Next we investigate the performance in a mmWave MIMO-OFDM system with $64$ subcarriers. Similar to that in the narrowband scenario, we first show the convergence for the proposed algorithms. Fig. \ref{convergenceWB} shows the averaged MSE over 1000 channel realizations as a function of the number of the outer iterations when $\mathrm{SNR}=-10\mathrm{dB}$ and $N_\mathrm{s}=N_\mathrm{RF}=2$ in the broadband scenario. For the VFD initialization method, we apply the result in \cite{Smapth:2001} to initialize the hybrid combiner. Like that in the narrowband scenario, the proposed VFD initialization method can reduce about 2-3 iterations for all the HBF algorithms, and thus greatly save the computational complexity, when compared with the random initialization method. %Particularly for the EVD-HBF algorithm, the VFD initialization method can also improve the final MSE performance.

We then illustrate the BER performance as a function of SNR for the four proposed algorithms (labeled with `MO', `EVD-LB', `EVD-UB', `OMP', respectively). Two system configurations are considered, where in Fig. \ref{ori_ber} we set $N_\mathrm{t}=N_\mathrm{r}=64, N_\mathrm{RF} = N_\mathrm{s}=2$, and in Fig. \ref{yuwei_ber} we set $N_\mathrm{t}=64, N_\mathrm{r}=32, N_\mathrm{RF}=4, N_\mathrm{s}=2$. For comparison, a full-digital scheme is straightforwardly extended from the result in \cite{Smapth:2001} and adopted as a performance benchmark. Comparing Fig. \ref{ori_ber} and Fig. \ref{yuwei_ber}, it can be seen that as the analog precoder or combiner is consistent over all the subcarriers in the broadband scenario, the gaps between the proposed HBF algorithms and the full-digital one are larger than those in the narrowband scenario, especially in the extreme case when $N_\mathrm{RF} = N_\mathrm{s}$. However, when more RF chains are available, as shown in Fig. \ref{yuwei_ber}, there are more optimization freedom in the HBF design and the gap to the full-digital one shrinks. Furthermore, comparing the four proposed HBF algorithms, the restriction on the feasible set of the analog beamformers in the OMP-HBF algorithm also leads to certain performance loss in the broadband scenario. As for both the EVD-UB-HBF and EVD-LB-HBF algorithms, the adoption of an upper or lower bound as a surrogate of the objective function and the phase extraction operation applied to obtain the final analog beamformers lead to some performance loss. However, the MO-HBF algorithm directly tackles the original problem without making approximations and therefore achieves much better performance than other algorithms. Finally, by comparing Fig. \ref{convergenceWB} and Fig. \ref{ori_ber}, it can be seen that for different HBF algorithms their difference in the average BER performance (plotted in a base 10 logarithmic scale) is more obvious than that in the corresponding MSE performance. To explain the phenomenon, we checked the performance of each channel realization and found that the average BER performance is mainly dominated by the channels in `bad' conditions and the average MSE performance is less affected by these channels.

\begin{figure}
 		\centering
 		\includegraphics[width=3.8in]{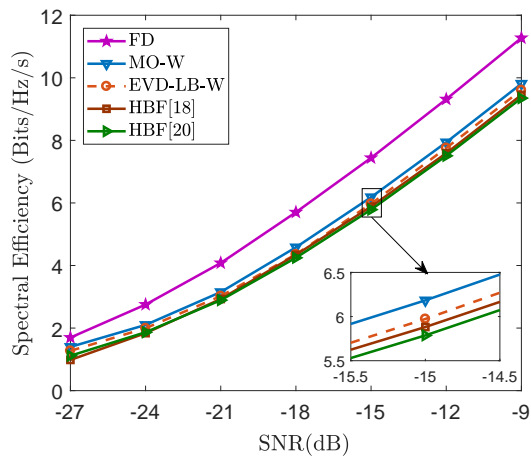}
 		\caption{Spectral efficiency v.s. SNR for different HBF algorithms when $N_\mathrm{RF} = N_\mathrm{s}=2$ in the broadband scenario.}
 		\label{ori_rate}
 	\end{figure}

\begin{figure}
 		\centering
 		\includegraphics[width=3.8in]{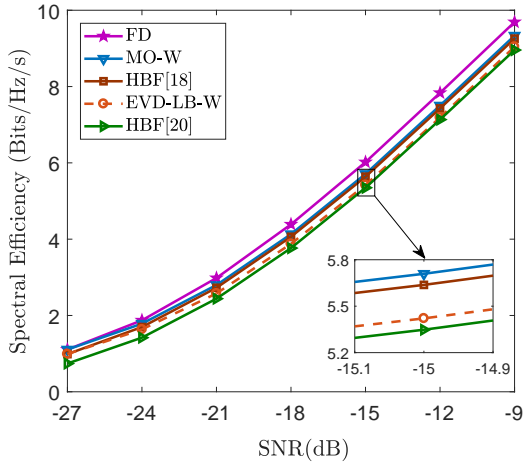}
 			\caption{Spectral efficiency v.s. SNR for different HBF algorithms when $N_\mathrm{t}=64, N_\mathrm{r}=32$, $N_\mathrm{RF}=4, N_\mathrm{s}=2$ in the broadband scenario.}
 		\label{yuwei_rate}
 		
 \end{figure}

Fig. \ref{ori_rate} and Fig. \ref{yuwei_rate}  show the spectral efficiency as a function of SNR in the above two broadband systems for the two proposed WMMSE HBF algorithms (labeled with `MO-W' and `EVD-LB-W') in Section \ref{sec:WMMSE}. For comparison, the performance of the full-digital beamforming algorithm and the two conventional HBF algorithms in \cite{ZhangJun2016} and \cite{YuWei2017} (labeled with 'HBF \cite{ZhangJun2016}' and 'HBF \cite{YuWei2017}') aiming at maximizing the spectral efficiency is also provided. It can be seen that, all the HBF algorithms perform close to each other with the little difference depending on the system configurations. The competitive performance of the proposed WMMSE HBF algorithms comes from the close connection between the WMMSE based formulation and the spectral efficiency objective, as well as the benefit of the alternating optimization of the transmitter and receiver beamformers. Furthermore, similar to that in Fig. \ref{yuwei_ber}, with more RF chains, the HBF performance becomes closer to the full-digital one.

% \begin{figure*}[!t]
% 	\centering
% 	\begin{center}
% 		\centerline{\subfigure[]{\includegraphics[width=2.5in]{yuwei_ber.eps}\label{yuwei_ber}} \hfil \subfigure[]{\includegraphics[width=2.5in]{yuwei_rate.eps}\label{yuwei_rate}}\hfil }
% 		\caption{Performance for different HBF algorithms when  $N_\mathrm{t}=64, N_\mathrm{r}= 32$ and $N_\mathrm{RF} = 4,  N_\mathrm{s}=2$ in the broadband scenario. (a)BER v.s. SNR. (b) The spectral efficiency v.s. SNR.} \label{fig:sim_WByuwei}
% 	\end{center}
% \end{figure*}

 %\begin{figure*}[!t]
 %	\centering
 %	\includegraphics[width=3.7in]{yuwei_ber.eps}%or 8.5cm
 %	\caption{BER v.s. SNR for different HBF algorithms when  $N_\mathrm{t}=64, N_\mathrm{r}= 32$ and $N_\mathrm{RF} = 4,  N_\mathrm{s}=2$ in the broadband scenario.}
 %	\label{yuwei_ber}
 %\end{figure*}

 %\begin{figure*}
 %	\centering
 %	\includegraphics[width=3.7in]{yuwei_rate.eps}
 %	\caption{ The spectral efficiency v.s. SNR for different HBF algorithms when  $N_\mathrm{t}=64, N_\mathrm{r}= 32$ and $N_\mathrm{RF} = 4,  N_\mathrm{s}=2$ in the broadband scenario.}
 %	\label{yuwei_rate}
 %\end{figure*}

 %To further verify the performance under different simulation parameters, Fig. \ref{yuwei_ber} and Fig. \ref{yuwei_rate} show the BER and sum-rate results with $N_\mathrm{t}=64, N_\mathrm{r}=32, N_\mathrm{s}=2,\NRF=4$. The figures suggest similar conclusions compared with Fig. \ref{yuwei_ber} and Fig. \ref{yuwei_rate}, which evidence that  our algorithms can guarantee performance in different simulation scenarios. One difference  should be mentioned that the HBF algorithm in \cite{YuWei2017} slightly outperforms the EVD-HBF algorithm. The main reason is that the performance of EVD-HBF relies on the approximation that  $\mathbf{V}_\mathrm{RF}^H\mathbf{V}_\mathrm{RF} \approx \mathbf{I}_{N_\mathrm{RF}}$ and $\mathbf{W}_\mathrm{RF}^H\mathbf{W}_\mathrm{RF} \approx \mathbf{I}_{N_\mathrm{RF}}$. However, the approximation gap becoming larger as $NRF$ increases, which leads to certain performance loss of EVD-HBF algorithm.

\begin{figure}[!t]
 	\centering
 	\includegraphics[width=3.8in]{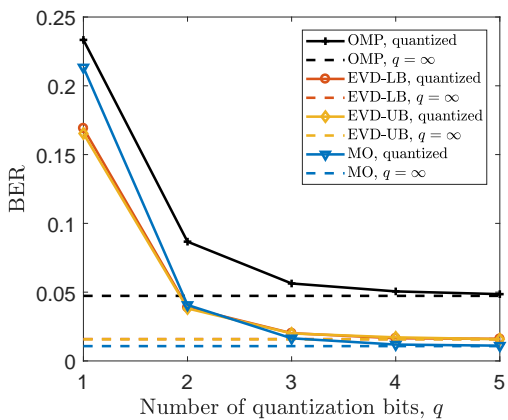}%or 8.5cm
 	\caption{BER v.s. the number of quantization bits $q$ when $N_\mathrm{RF} = N_\mathrm{s}=2$, $\mathrm{SNR}=-16\mathrm{dB}$ in the broadband scenario.}
 	\label{fig:quantized}
\end{figure}

Finally, considering that practical phase shifters have limited resolution, we uniformly quantize the analog beamforming coefficients with a limited number of bits and provide in Fig. \ref{fig:quantized}  the BER performance of different HBF algorithms as a function of the number of quantization bits, $q$, when $N_\mathrm{RF} = N_\mathrm{s}=2$, $\mathrm{SNR}=-16\mathrm{dB}$. It can be seen that the performance loss due to the finite resolution decreases as $q$ increases and becomes negligible when $q\geq 5$. While a simple quantization method is adopted here,  the investigation of the HBF algorithms with more sophisticated quantization methods is of great interest in future work.\\

\section{Conclusions}\label{sec:conclusion}
In this paper, we investigated the HBF optimization for broadband mmWave MIMO communication systems. Instead of maximizing the spectral efficiency as in most existing works, we took the MSE as a performance metric to characterize the transmission reliability. To directly minimize MSE, several efficient algorithms were proposed based on the principle of alternating optimization. The MMSE based HBF design was also extended to the WMMSE one and further used to solve the spectral efficiency maximization problem. Simulation results showed that the BER and spectral efficiency performance of the proposed MO-HBF algorithm approaches the full-digital beamforming with much fewer RF chains, while other low-complexity algorithms balance the system performance and computational complexity. For future research directions, the HBF designs with other objectives such as BER minimization and with finite resolution phase shifters are also of great interests to be investigated.% the extension to the multi-user scenario requires more effort, and channel estimation and finite resolution phase shifters are also of great interests to be investigated.

\section{ACKNOWLEDGMENT}
The authors would like to thank Dr. Xianghao Yu at Friedrich-Alexander-Universit\"at Erlangen-N\"urnberg for his kind help in this work. In addition, we also thank the students Mr. Tianyi Lu
and Ms. Anqi Jiang at Fudan University for their help in the
proof of Lemma 3.

\end{document}